\documentclass[]{jkas} 

\def\year{2025} 
\def\volume{56} 
\def\issue{1} 
\def\beginpage{1} 
\def\received{---} 
\def\accepted{---} 
\def\published{---} 
\date{Received \received; Accepted \accepted; Published \published}

\newcommand\ion[2]{{#1}\,\textsc{#2}} 

\title{%
The Identification of Asymmetric Barred Galaxies in Illustris TNG-50 
}

\author[1]{Junwoo Jung}{}
\author[2,$\star$]{Minjin Kim}{0000-0002-3560-0781}
\author[1]{Taehyun Kim}{0000-0002-5857-5136}

\affil[1]{Department of Astronomy and Atmospheric Sciences, Kyungpook National University, Daegu 41566, Republic of Korea}
\affil[2]{Department of Astronomy, Yonsei University, 50 Yonsei-ro, Seoul 03722, Republic of Korea}

\def\corrauthor{%
M. Kim, \email{mkim.astro@yonsei.ac.kr}
}

\def\runningauthor{%
Jung et al.
}

\def\runningtitle{%
Asymmetric-barred Galaxies in TNG50-1
}

\def\keywords{%
astronomy --- astrophysics --- journals: individual: JKAS
}

\def\abstracttext{
Most barred galaxies exhibit symmetric structures. However, recent studies have shown that a subset of barred galaxies exhibit lopsided morphologies. To quantify their occurrence and investigate their physical origins, we analyze barred galaxies in the IllustrisTNG TNG50 simulation. We select 519 clearly barred galaxies in their stellar mass maps out of 770 barred galaxies from the TNG50-1 catalog. We classify the bar morphologies into four subgroups - `Lopsided', `Perturbed', `Symmetric', and `Indeterminate' - and perform a comparative analysis of their physical properties. We find that galaxies hosting asymmetric bars (`Lopsided' and `Perturbed') tend to have higher gas densities around the bar region, enhanced star formation activity, and more recent bar-formation epochs than galaxies with symmetric bars. However, the factor that most consistently distinguishes the four subgroups is the stellar mass distribution of the host galaxy, and there appears to be no physical correlation with bar size. These findings suggest that asymmetric bars form preferentially in less massive galaxies and may evolve into symmetric bars over time through secular processes. However, this conclusion should be considered with caution, as the fraction of asymmetric bars in the TNG50 simulation is systematically higher than that observed in the local universe. 
}

\begin{document}
\jkashead 


\section{Introduction} \label{sec:intro}
Barred galaxies are common among disk galaxies in the local Universe. While the bar fraction varies with stellar mass, color, and morphological type, it has been reported to be up to $\sim 65\%$ \cite[e.g.,][]{eskridge_2000, menedez_2007, sheth_2008, masters_2011, buta_2015, wang_2025}. As bars are non-axisymmetric structures, they drive the transfer of angular momentum, causing gas to flow inward and accumulate in the central regions of galaxies \cite[][]{athanassoula_1992a, sellwood_1993, sakamoto_1999, sheth_2000, jogee_2005, kuno_2007, kim_2012, sormani_2015, fragkoudi_2016, seo_2019}. This inflow triggers a burst of star formation in the central regions \cite[e.g.,][]{hunt_1999, ho_1997, jogee_2005, ellison_2011} and even fuels AGN \cite[][]{ho_1997, cheung_2013}. Barred galaxies have been found to exhibit a higher incidence of AGN activity \cite[e.g.,][]{simkin_1980, galloway_2015, alonso_2018}; however, after controlling for sample properties, no significant difference is observed \cite[e.g.,][]{lee_2012, cheung_2015, cisternas_2015}, indicating that the interplay between the bars and AGNs is complex \cite[e.g.,][]{oh_2012, zee_2023}. Ample gas accumulated in the central regions of barred galaxies naturally leads to the formation of nuclear disks \cite[e.g.,][]{gadotti_2020, desafreitas_2023}, and can even give rise to secondary structures such as nuclear bars \cite[e.g.,][]{delorenzo_2019} or nuclear spirals \cite[e.g.,][]{vandeven_2010ApJ...723..767V}. Consequently, as these nuclear structures develop and reshape the central dynamics, bars are thought to drive the formation of pseudo-bulges \cite[][]{kormendy_2004, kormendy_2005, gadotti_2009}. 
The influence of bars extends even beyond the nuclear regions, actively reshaping the global mass distribution of galaxies. This process leads to the emergence of various features, including inner and outer rings, spiral arms, and dark gaps \citep{buta_2009a, salo_2010, buta_2017b, james_2018, ghosh_2024b, kim_2025b}.

Barred galaxies exhibit largely symmetric bar structures. However, lopsided bars are frequently observed in Sd and later Hubble types \cite[e.g.,][]{odewahn_1996, deswardt_2015, kruk_2017}. Off-centered bars show a clear deviation from axial symmetry, in which the photometric centers of the disk and the bar do not coincide. The Large Magellanic Cloud (LMC) is a canonical example and is often regarded as the prototype of Magellanic-type galaxies. Such offsets have been attributed to interactions with neighboring galaxies \cite[][]{athanassoula_1996, berentzen_2003, pardy_2016}, perturbations induced by the dark-matter (sub)halo (\citealp{bekki_2009}), or the presence of a lopsided halo potential.
\citet{kruk_2017} identified 271 galaxies hosting offset bars, characterized by a 0.2--2.5 kpc displacement between the photometric centers of the disk and bar. These systems exhibit masses, colors, and degrees of offset comparable to those of the LMC. Interestingly, the presence of an offset bar does not strongly correlate with the distance to the nearest companion, and many isolated galaxies also exhibit such offsets.

In addition to off-centered systems, another class of lopsided barred galaxies has been identified in which the bar itself is asymmetric, with one side thinner, thicker, or more curved than the other \cite[][]{lokas_2021, cuomo_2022, sanchezmartin_2023}. Simulations by \citet{colin_1989} showed that some off-centered barred galaxies can develop such asymmetric bars, particularly when the disk center is displaced along the bar axis (their Fig. 1g and h). Using the TNG100 simulation, \citet{lokas_2021} identified six galaxies with asymmetric bars and demonstrated that these features can persist for a few Gyrs. To investigate their origin, they performed controlled simulations including interactions with a companion and configurations in which the disk is offset relative to the dark matter halo. Both scenarios produced asymmetric bars. Notably, in the interaction-driven cases, the asymmetry disappeared after bar buckling as stellar orbits were reconfigured, whereas in the halo–disk offset case, the asymmetry persisted for a while but faded prior to buckling. However, previous studies on bar asymmetry or lopsidedness have largely been restricted to a limited number of galaxies \cite[e.g.,][]{lokas_2021, cuomo_2022}. Consequently, the physical origin of the asymmetric bar is not yet fully established, making it essential to statistically explore the properties of asymmetric bars using a significantly larger sample of galaxies.

The IllustrisTNG project is a series of extensive magnetohydrodynamical cosmological simulations that model galaxy formation and evolution in a self-consistent manner \cite[][]{Pillepich2018, nelson_2019}. As a result, several studies utilizing TNG simulations have examined the formation, evolution, demographics, and properties of barred galaxies \cite[e.g.,][]{Peschken2019, RosasGuevara2020, Zhao2020, zhou_2020, RosasGuevara2022, semczuk_2024}. In particular, TNG50 provides high-resolution simulated galaxies, allowing the structure of bars to be well resolved, and offers detailed imaging analysis that enables a quantitative assessment of bar strength \cite[][]{pillepich_2019, nelson_2019, zana_2022}. Therefore, it presents a valuable opportunity to identify asymmetric bars and their physical origins based on cosmological simulations.

In this paper, we investigate the properties of galaxies exhibiting bar asymmetry, with the goal of probing the physical mechanisms that give rise to such lopsided structures. The paper is organized as follows. We describe the data and sample selection in Section \ref{sec:data}. 
We classified galaxies into asymmetric and symmetric bars in Section \ref{sec:classification}. We analyze the gas morphology around the bars and compare their stellar masses and star formation rates to the local star-forming main sequence in Section \ref{sec:results} and discuss its physical origin in Section \ref{sec:discussion}. Throughout the paper, we adopt the cosmological parameters ($H_0=67.74$ km s$^{-1}$ Mpc$^{-1}$, $\Omega_m=0.3089$, and $\Omega_\lambda=0.6911$) from \citet{planck_2016}. 

\begin{figure}[t]
\centering
\includegraphics[width=0.48\textwidth]{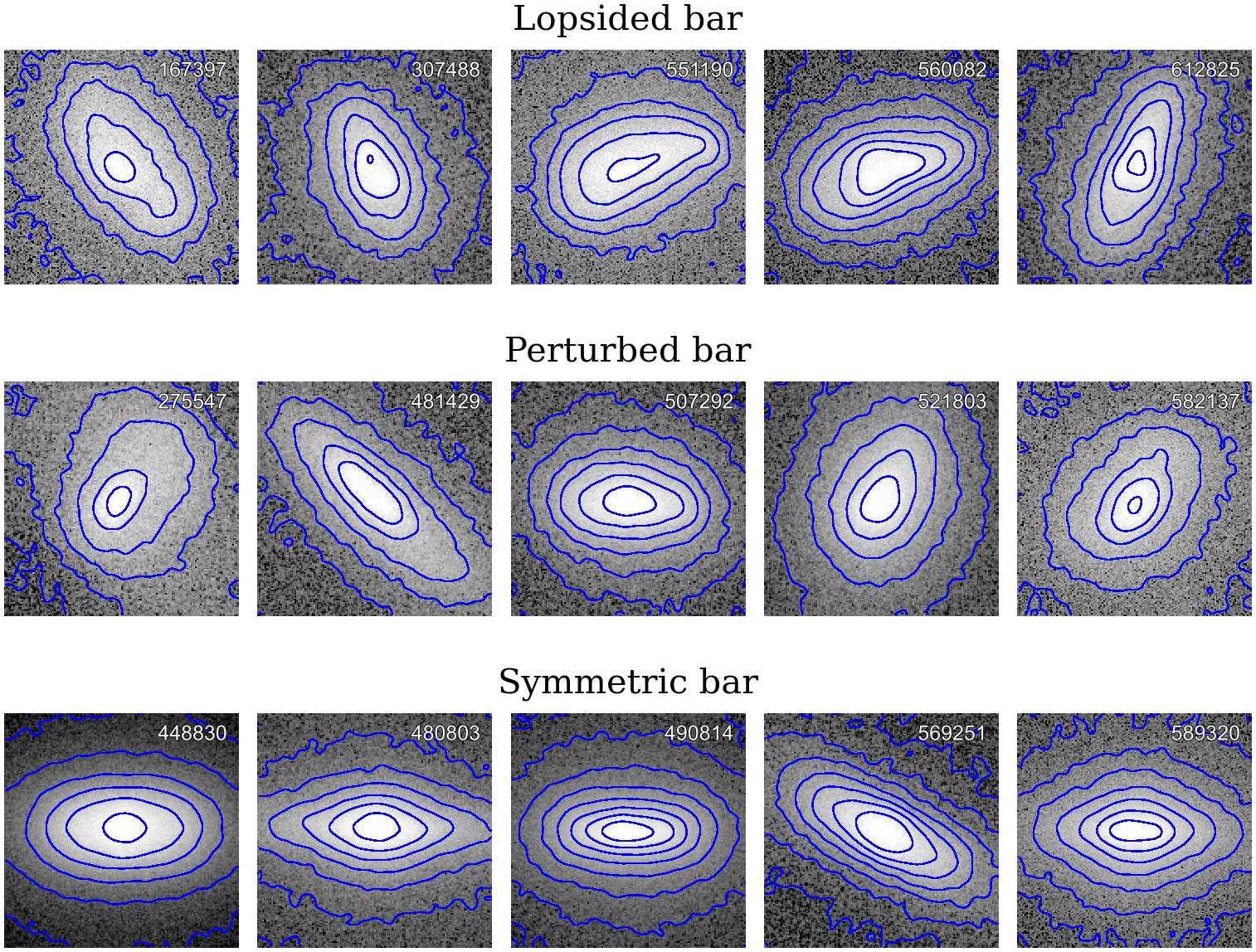}
\caption{Stellar mass contour maps for various subsamples. From top to bottom, `Lopsided', `Perturbed', and `Symmetric' barred galaxies are shown, respectively, as identified through visual inspection. 
Each image covers an area of $2.25 \times 2.25$ kpc.
}
\vspace{0mm} 
\label{fig:mass_maps}
\end{figure}

\section{Simulation Data and Sample} \label{sec:data}
We used mock data from TNG50 of the IllustrisTNG project, a cosmological and magnetohydrodynamical simulation performed within a $\sim 50$ cMpc-sized box. Because TNG50 offers the highest resolution within the IllustrisTNG suite, it enables a detailed investigation of the structural and physical properties of barred galaxies \citep{nelson_2019}. We initially selected a sample of barred galaxies at $z = 0$ (i.e., snapshot 99 of the TNG50 simulation), as described in \citet{zana_2022}. 

\begin{figure}[t]
\centering
\includegraphics[width=0.48\textwidth]{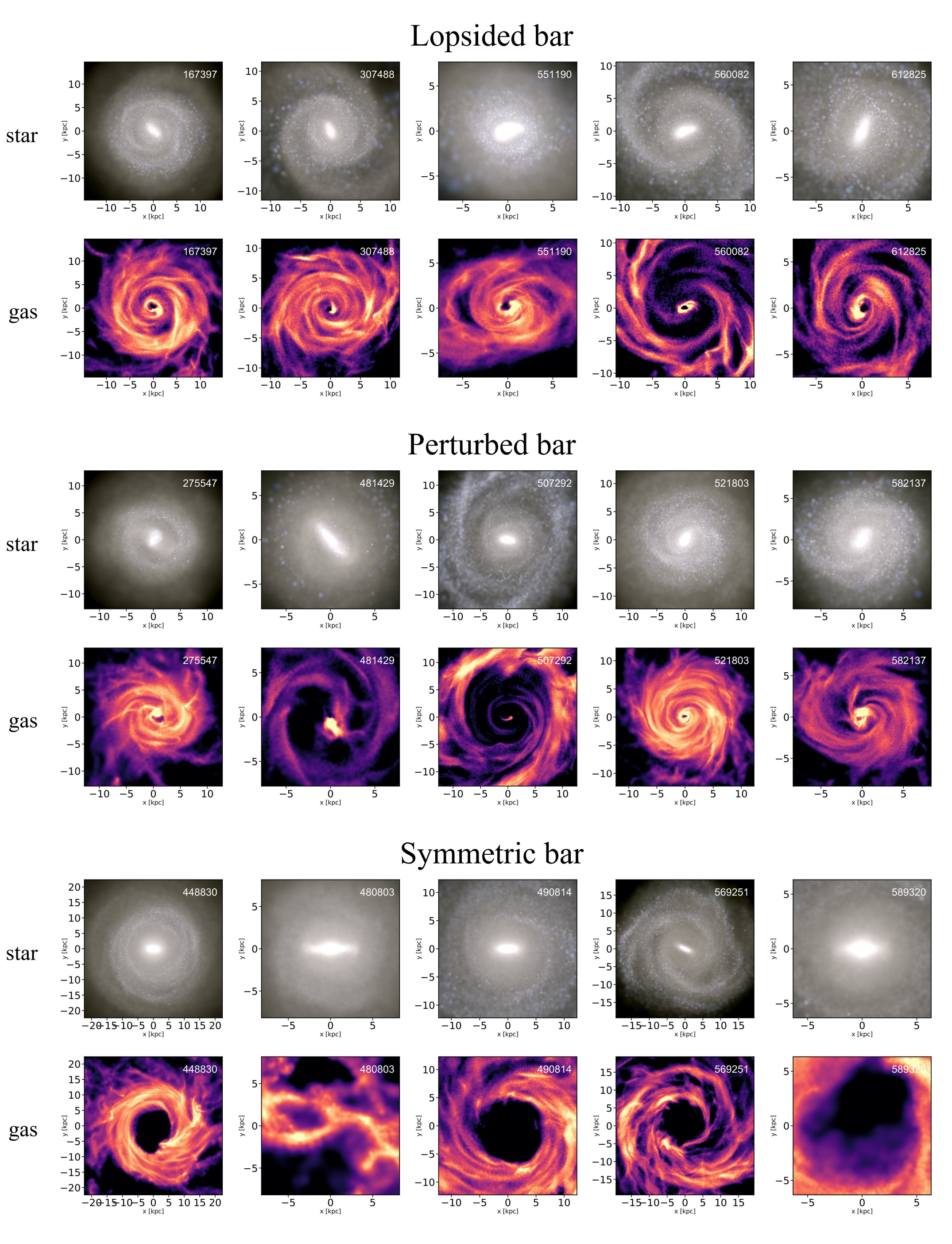}
\caption{Stellar light and gas density distributions for 15 representative barred galaxies. The stellar light (JWST/NIRCam F070W, F115W, and F200W) is shown in gray scales, while gas density maps are shown in red stretches. The galaxies are organized by morphological classification: the top two rows display `Lopsided' bars, the middle two rows show `Perturbed' bars, and the bottom two rows show `Symmetric' bars. All classifications were determined via visual inspection.}
\vspace{0mm} 
\label{fig:light_gas}
\end{figure}

To investigate the non-axisymmetric structures of galaxies, Fourier decomposition of the two-dimensional surface brightness distribution of galaxies is often employed. The azimuthal profile of galaxies at a given radius $r$ is expressed as
\begin{equation} \label{eq:fourier}
 I(r,\theta) = {a_0(r)} + \sum_{m=0}^{\infty} \left[ a_m(r)\cos(m\theta) + b_m(r)\sin(m\theta) \right],\\
\end{equation}
where, 
\begin{gather*}
a_m(r) = \frac{1}{\pi} \int_{0}^{2\pi} I(r,\theta)\cos(m\theta)\, d\theta , \\
b_m(r) = \frac{1}{\pi} \int_{0}^{2\pi} I(r,\theta)\sin(m\theta)\, d\theta, 
\end{gather*}
and the Fourier amplitude of m-th component can be defined as follows,
\begin{equation} 
A_m(r) = \sqrt{a_m(r)^2 + b_m(r)^2},
\end{equation}
where $\theta_m(r)$ denotes its phase angle. This formalism provides an objective and quantitative way to characterize non-axisymmetric features of galaxies. In this decomposition, the $m=1$ mode captures global lopsidedness of galaxies, the $m=2$ mode traces stellar bars or spiral arms; and higher-order modes ($m \ge 3$) reveal more complex structures, such as multi-armed spirals or tidal distortions \cite[e.g.,][]{yu_2018}. The phase terms further provide information on the orientation and spatial coherence of these components. In particular, stellar bars produce a strong and coherent bisymmetric ($m=2$) component whose phase remains nearly constant from the galaxy center out to the end of the bar. As a result, a bar exhibits a pronounced $m=2$ Fourier amplitude with a nearly constant $m=2$ phase over a contiguous radial interval. In contrast, spiral arms, tidal disturbances typically exhibit varying phases.

To describe the bar strength, the $m=2$ amplitude $A_2(r)$ is considered. The maximum value of $A_2(r)$ within the bar-dominated region, hereafter referred to as $A_2$, provides a direct measure of the bar strength relative to the underlying axisymmetric disk. The bar strength can be defined as 
\begin{equation} \label{eq:bar_strength}
    A_{\mathrm{2}} = \max\left(\frac{A_2(r)}{A_0(r)}\right)
\end{equation}
which has been widely adopted \cite[e.g.,][]{ohta_1990, laurikainen_2002, buta_2005, aguerri_2009, athanassoula_2013, lee_2020}. \citet{zana_2022} identified barred galaxies using $A_2$. Specifically, galaxies with a maximum $A_2$ greater than 0.1 were classified as barred. This criterion resulted in a sample of 770 galaxies in the TNG50 simulation, which are presented in TNG50-1 catalog. We find that the $A_2$ parameter alone is insufficient to distinguish between barred and non-barred galaxies, as their distributions in $A_2$ space frequently overlap in observations (Kim et al., in prep.). Furthermore, the initial selection criterion for bar classification ($A_2 > 0.1$) potentially includes weak bars, complicating the robust identification of asymmetric bars from the parent sample. Consequently, we performed a visual inspection of galaxies exhibiting clear bar structures in the stellar mass maps. This yielded a final sample of 519 targets, of which approximately $87\%$ satisfy the criteria for a strong bar ($A_2 > 0.2$).  

\begin{figure*}[t]
\centering
\includegraphics[width=0.98\textwidth]{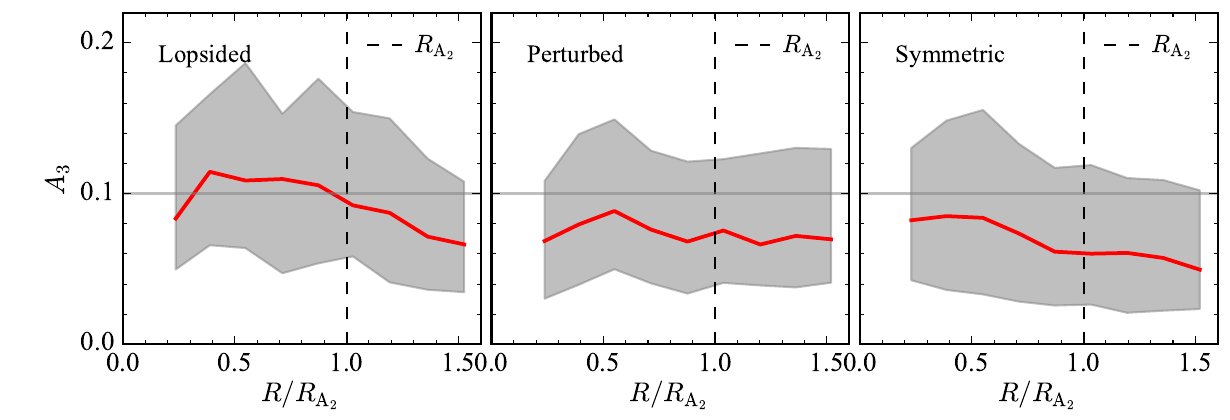}
\caption{Ellipse fitting results ($A_3$ coefficient) of galaxies for different subgroups performed using \texttt{Photutils}. The red solid line denotes the median value, and the shaded region represents the 16th and 84th percentiles of the subsamples. The semimajor axis is normalized by the bar length ($R_{A_{2}}$).
Vertical dashed lines indicate the bar radius. The horizontal gray lines denote the $A_3 = 0.1$ threshold, a standard criterion for identifying bar lopsidedness. To ensure data quality in the $A_3$ radial profiles, we consider only measurements meeting the reliability criterion of $\sigma_{A_{3}} < 0.1$.}
\vspace{0mm}
\label{fig:ellipse}
\end{figure*}

\section{Bar Classification} \label{sec:classification}
\subsection{Parametric Classification}
The amplitude of the $m=3$ Fourier mode ($A_3$; e.g., \citealp{lokas_2021}) has been adopted to identify asymmetric bars by quantifying the lopsidedness of the feature. However, the relevance of this parameter-based classification can be sensitive to spatial resolution, nuclear spirals, and flocculent features within the galaxies. To assess the reliability of this method for our sample, we calculated the $A_3$ amplitude across a radial range from 0.29 kpc to the bar radius, $R_{\rm A_2}$. The bar radius, $R_{\rm A_2}$, was adopted from \citet{zana_2022}, who determined it using the $A_2$ parameter, where $A_2(r)$ reaches its peak. We decided to use the ellipse fitting results only beyond the inner radius of 0.29 kpc because the $Z=0$ Plummer-equivalent gravitational softening for the collisionless component in TNG50-1 was set to $0.29 \text{ kpc}$ in the simulation, thus potentially introducing biased $A_3$ measurements within this radius. The ellipse fitting was applied to the stellar density map using the Python package \texttt{photutils}. We chose to use the stellar density map instead of the SKIRT images \cite[][]{camps_2015, baes_2011} to make the galaxy appear face-on, which is crucial for robustly estimating the asymmetry of the bar. During the fit, we fixed the central position as given by the simulation to improve the reliability of the $A_3$ parameter.  

We explored two approaches to define a proxy for asymmetry: adopting either the maximum or the mean value of $A_3$ within the bar radius. Subsequently, we applied selection criteria ($A_3 = 0.2$) to classify bars as asymmetric or symmetric. However, our visual assessments of the stellar mass maps revealed that $A_3$-based classification is often misleading and results in ambiguous cases. Therefore, we concluded that the parametric classification may not be suitable for our purpose.

\subsection{Visual Classification}
We adopted visual inspection instead of the $A_3$-based method to identify asymmetric barred galaxies. Based on visual inspection of face-on stellar mass maps, we classified the barred galaxies into four morphological categories according to their degree of asymmetry: `Lopsided', `Perturbed', `Symmetric' and `Indeterminate'.

We classified a `Lopsided' bar when the bar itself shows a clear, large-scale asymmetry between the two sides. This feature is recognized as a noticeable difference in bar length and/or thickness on opposite sides of the center, or as a strong one-sided curvature (`banana-like' or `footprint-like') in the main body of the bar or its outer extensions. This classification yielded 29 objects out of 519 barred galaxies. The stellar mass maps of all `Lopsided' bars, alongside their contours, are shown in Figure \ref{fig:mass_maps}. On the other hand, a subset of the sample exhibits subtle asymmetry localized to the central region rather than extending to the entire bar. We classified these features as `Perturbed' when the bar is clearly present but shows only mild departures from bilateral symmetry. These departures are typically localized (e.g., small tilts, slight distortions, or modest one-sided irregularities) and do not produce a strongly one-sided, globally extended bar. This class corresponds to weakly asymmetric bars, in which the asymmetry is subtle compared to the `Lopsided' class, resulting in 58 objects. We emphasize that the classification of `Lopsided' and  `Perturbed' bars is based solely on the morphology of the bar region, as identified from the stellar mass contours, and excludes asymmetries that are confined to the disk.

To serve as counterparts to the `Lopsided' bars, we selected a control sample of `Symmetric' barred galaxies when neither side of the bar shows a systematically longer, thicker, or more strongly curved structure than the other, within the limits of visual uncertainty under a consistent image stretch and contouring scheme, yielding 283 objects. Minor small-scale irregularities (e.g., patchy features plausibly associated with spiral arm overlap or clumpiness) are allowed, provided they do not create a coherent, large-scale one-sided bar morphology. Figure \ref{fig:light_gas} shows the visual differences among the `Lopsided' class, `Perturbed' class, and `Symmetric' class in the stellar light map and gas density map. 

\begin{figure}[t]
\centering
\includegraphics[width=0.49\textwidth]{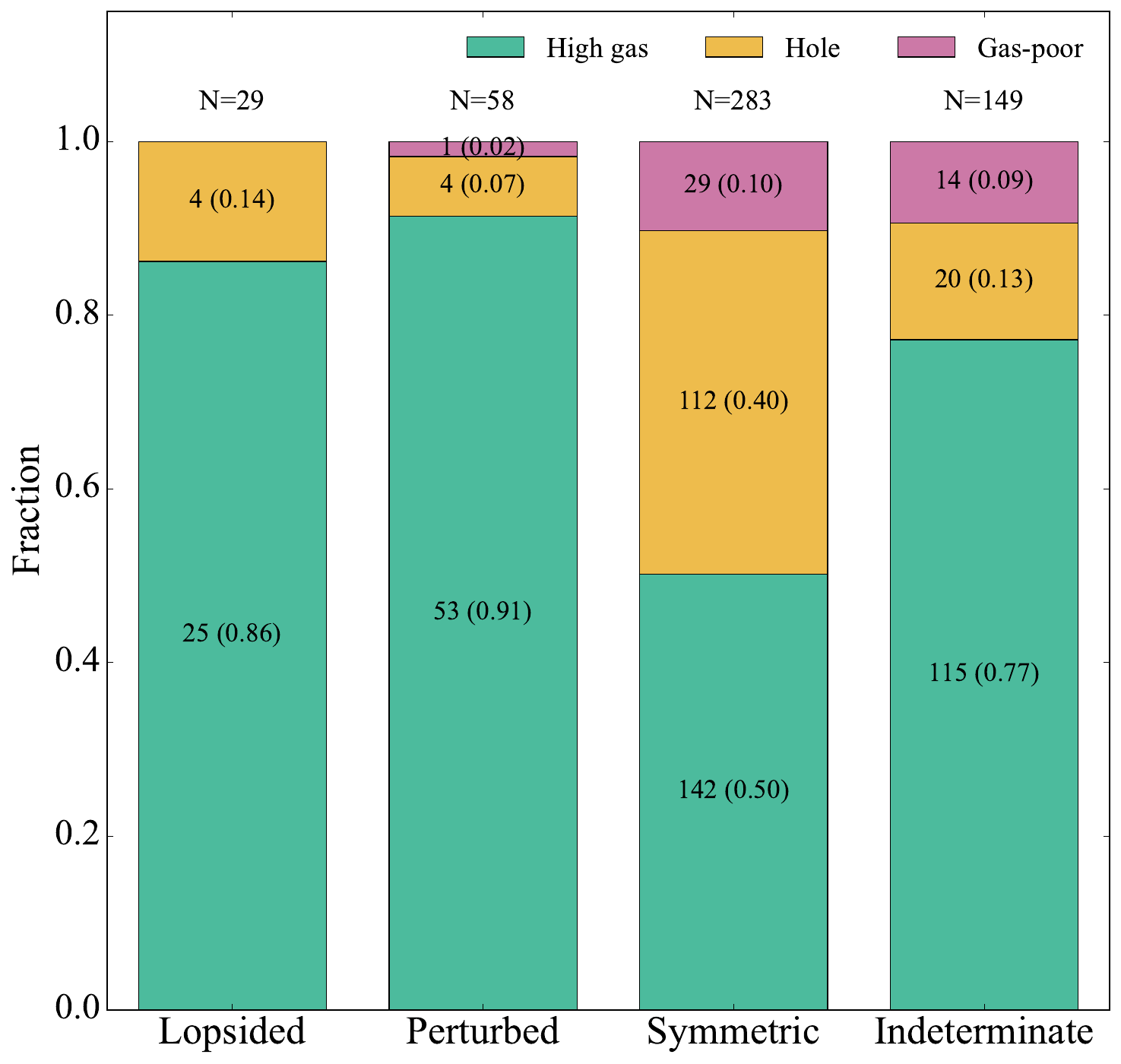}
\caption{Distribution of gas morphologies within each subsample. The fractions are calculated based on the visual classification criteria defined in Section \ref{sec:results}.}
\vspace{0mm} 
\label{fig:gas_fractions}
\end{figure}

Figure \ref{fig:ellipse} shows the median radial profiles of ellipticity and $A_{3}$ for `Lopsided', `Perturbed', and `Symmetric' bars, obtained from ellipse fitting to the mass maps. The radial behavior of the $A_3$ parameter reveals marginal differences between the three subsamples. On average, $A_3$ values for `Lopsided' bars exceed the 0.1 threshold within the bar radius, albeit with significant scatter. Conversely, `Symmetric' bars typically maintain $A_3$ values below 0.1. `Perturbed' bars exhibit moderate $A_3$ amplitudes that are systematically lower than those of the `Lopsided' subsample, potentially indicating a weak asymmetry signal. While $A_3$ serves as a useful proxy for bar lopsidedness \citep[e.g.,][]{lokas_2021}, these results suggest it may not provide a definitive classification on its own.

\begin{figure}[t]
\centering
\includegraphics[width=0.49\textwidth]{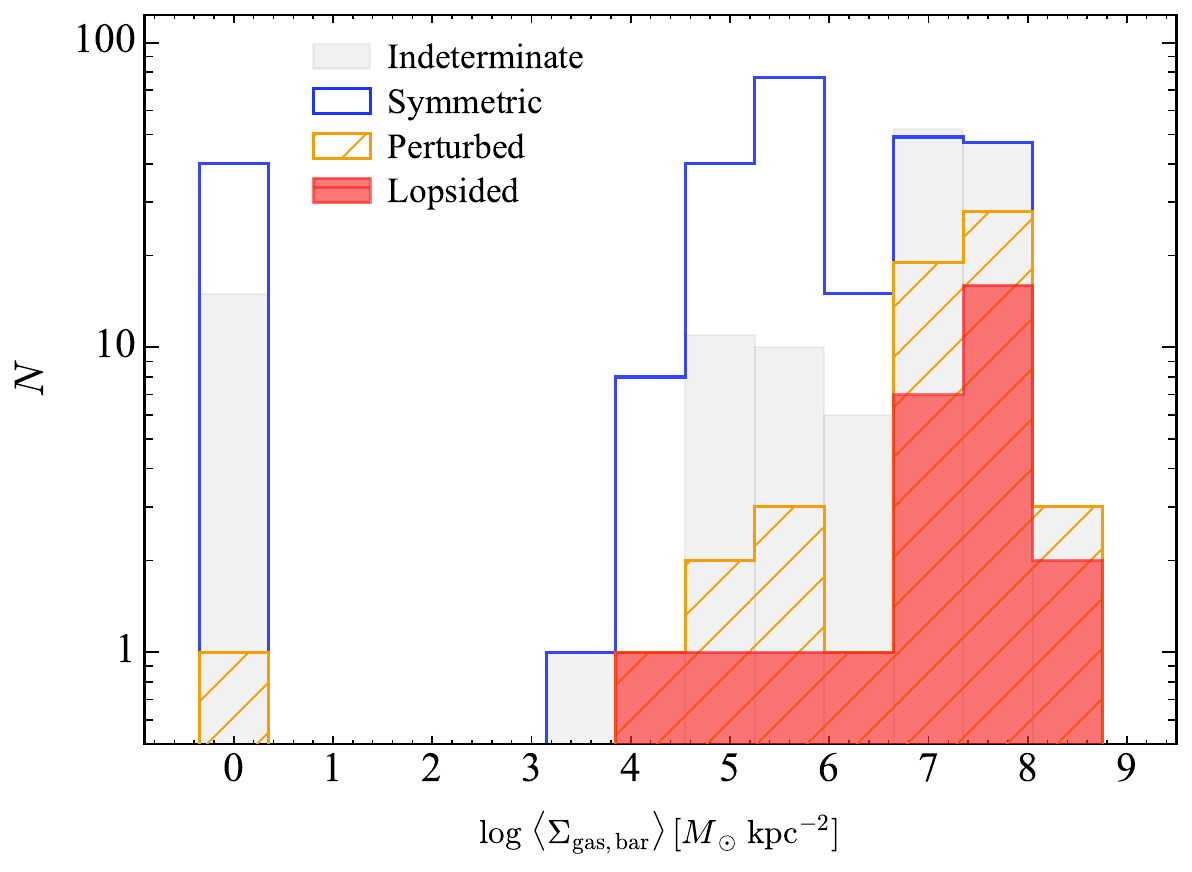}
\caption{Distribution of the gas surface density within the bar radius ($\Sigma_{\rm gas, bar}$) across the four galaxy subsamples. Galaxies with no detectable gas are plotted at a nominal value of $\log(\Sigma_{\rm gas,\ bar}) = 0$. }
\vspace{0mm} 
\label{fig:gasdensity}
\end{figure}

Finally, our visual inspection reveals that a subset of 149 galaxies cannot be clearly classified into either the asymmetric or symmetric bar classes. This occurs primarily when (1) the surface brightness or contrast of the bar is too low for a reliable assessment, or (2) the bar/disk (or bar/spiral) separation is ambiguous, such that apparent asymmetries could plausibly arise from surrounding structures rather than the bar itself. To maintain the integrity of our classification scheme, we classified the remaining bars in this analysis as `Indeterminate', which can not be processed by robust morphological judgment of bar symmetry. These four categories provide a morphology-driven framework for separating strongly asymmetric bars (`Lopsided'), slightly asymmetric bars (`Perturbed'), visually regular bars (`Symmetric'), and bars of which morphology cannot be reliably constrained (`Indeterminate'). In summary, for our barred sample, $5.6\%$, $11.2\%$, $54.5\%$, and $28.7\%$ are classified as `Lopsided', `Perturbed', `Symmetric', and `Indeterminate', respectively. 

The visual inspection was performed independently by two authors (J.J. and T.K.). The resulting classifications were consistent for 97\% of the sample. Discrepancies primarily occurred in distinguishing `Lopsided' from `Perturbed' morphologies, where the distinction can be inherently ambiguous. However, both subgroups exhibit similar physical properties, including stellar mass, gas abundance, and star formation rate (SFR). Therefore, this ambiguity does not alter the primary conclusions of this study. Final classifications were determined through iterative discussion and consensus between the two authors.

\begin{figure}[t]
\centering
\includegraphics[width=0.49\textwidth]{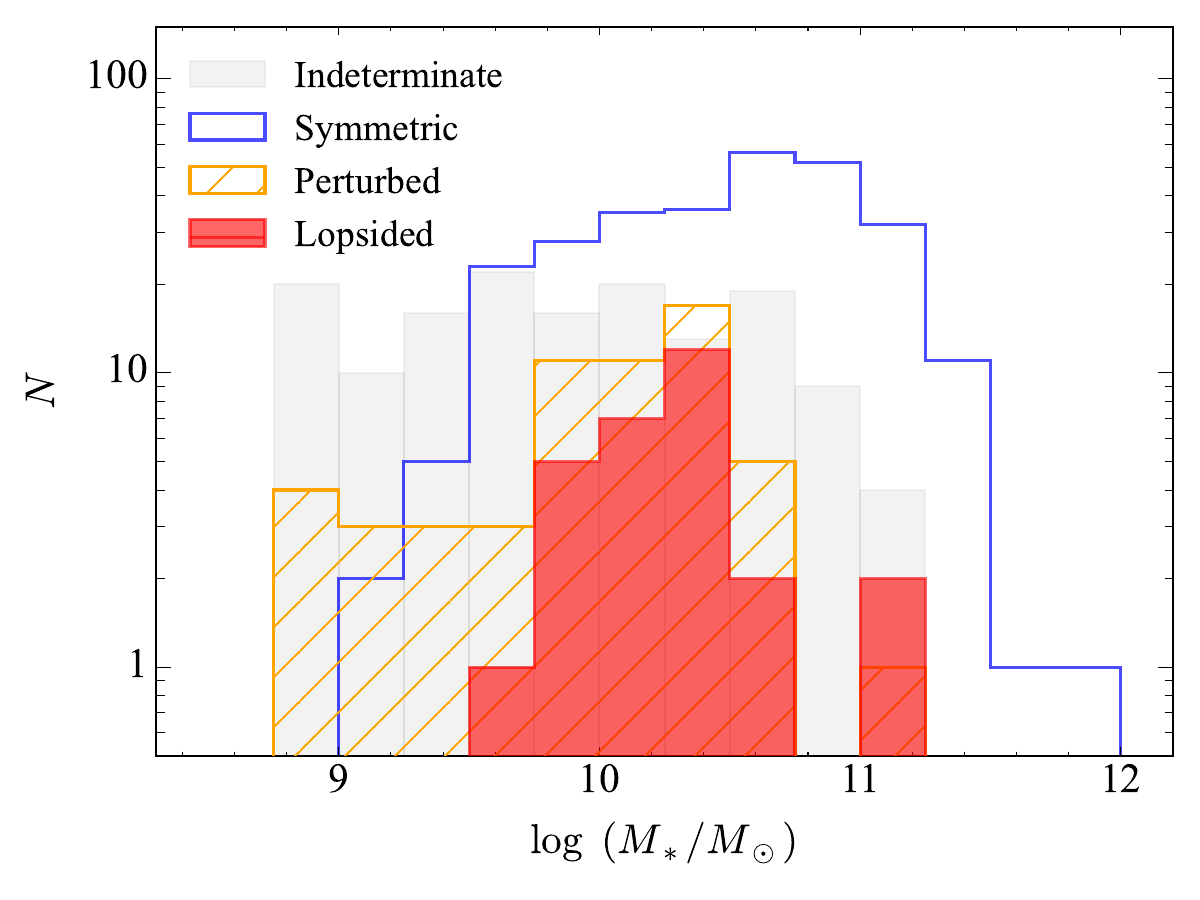}
\caption{Stellar mass distribution of four subsamples categorized by the bar disturbance levels. The histograms represent `Lopsided' (highly asymmetric; red-filled), `Perturbed' (slightly asymmetric; orange-hatched), `Symmetric' (blue), and `Indeterminate' (gray) bars.}
\vspace{0mm} 
\label{fig:smass}
\end{figure}

\begin{figure*}[t]
\centering
\includegraphics[width=1\textwidth]{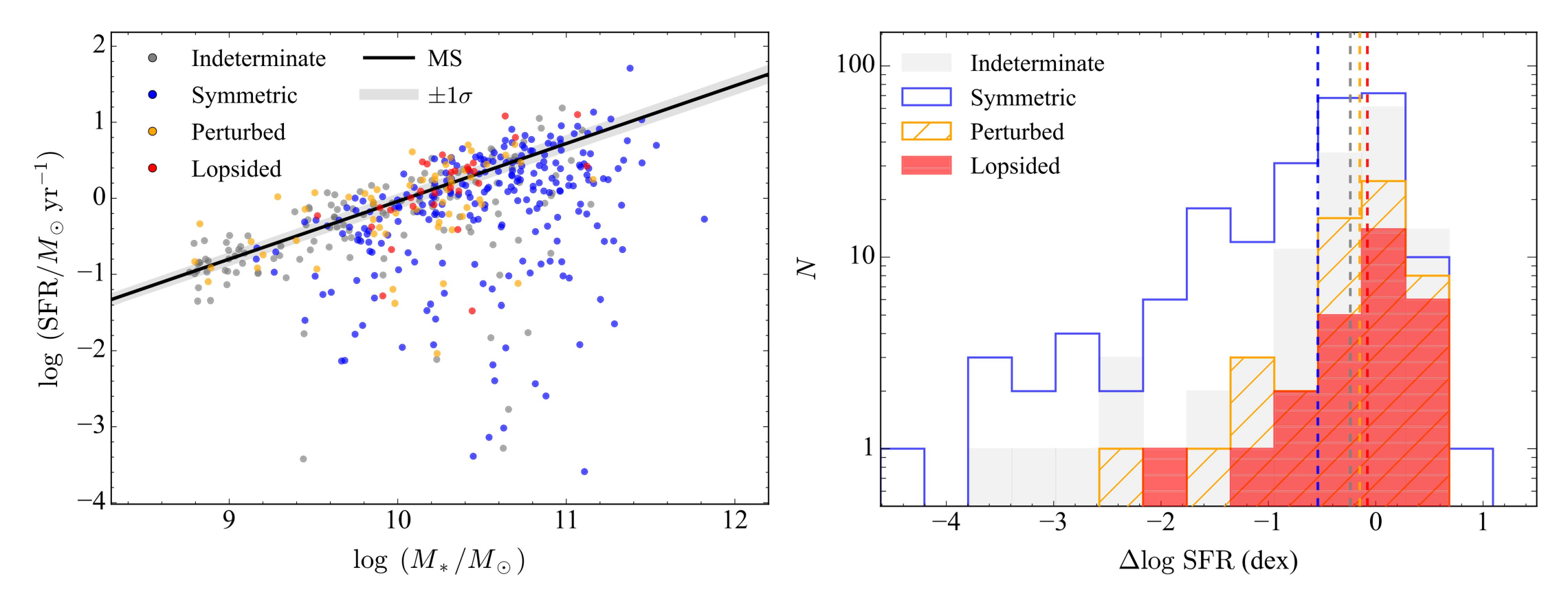}
\caption{Star formation properties of the sample subgroups. Left: Star formation rate (SFR) as a function of stellar mass ($M_{\ast}$) for the four subgroups. The solid black line represents the local star formation main sequence (SFMS) from \citet{renzini_2015}. Right: Distribution of the SFR offset ($\Delta$SFR) from the local SFMS. Vertical dashed lines indicate the mean offset for each respective subgroup.}
\vspace{0mm} 
\label{fig:sfms}
\end{figure*}

\section{Result} \label{sec:results}
\subsection{Gas Contents}
To understand the physical origin of the asymmetric bar, we first examined the gas distribution of sample galaxies. Bar properties are closely related to the gas content of their host galaxies \cite[e.g.,][]{athanassoula_2013}. Strong bars are found to be generally gas-poor (Masters et al. 2012). It has been suggested that gas inside the corotation radius flows inward \cite[][]{combes_1994, kormendy_2004, athanassoula_2013}. Eventually, this would produce a central hole in the gas distribution. \citet{newnham_2020} found that galaxies with an \ion{H}{I} hole tend to exhibit relatively long bars and occupy regions marginally below the SFMS. They argued that \ion{H}{I} holes are linked to bar-driven gas depletion, indicating dynamically old bars. In contrast, galaxies without an \ion{H}{I} hole typically host shorter bars. As it has been proposed that gas-rich galaxies experience delayed bar formation \cite[][]{berentzen_2007, athanassoula_2013}, these galaxies may not have developed an \ion{H}{I} hole yet because their bars are dynamically young.

We constructed face-on gas surface density maps using gravitationally bound gas cells (PartType 0) associated with each subhalo in the TNG50-1 simulation.
This map displays the line-of-sight integrated gas mass surface density, $\Sigma_{\rm gas}$, in units of $M_\odot\,\mathrm{kpc}^{-2}$ with a logarithmic stretch. As in the stellar light maps, the field of view is set to five times the stellar half-mass radius of each galaxy.   

Visual inspection of these gas maps for our barred galaxies reveals a striking diversity in the distribution of cold gas around the bar region, ranging from strong central concentrations to ring- or cavity-like structures and nearly gas-free disks. Motivated by this diversity and previous work showing that bars can funnel gas into the central kiloparsec or redistribute it into rings and depleted regions \cite[e.g.,][]{sakamoto_1999, sheth_2000, sheth_2005, jogee_2005, athanassoula_2009, sormani_2015, waters_2024}, we classified the gas morphology into three representative types: high gas density near the bar, central hole (or cavity) structures, and gas-poor disks. The high gas density near the bar is characterized by enhanced surface density in the vicinity of the stellar bar and is shown in the second row of Figure \ref{fig:light_gas}. The central hole structures represent a ring- or cavity-like gas distribution, as in the bottom of Figure \ref{fig:light_gas}. The gas-poor disk exhibits effectively undetectable gas content, with a negligible gas mass and star formation rate in the TNG50 simulation.

Figure \ref{fig:gas_fractions} illustrates the distribution of gas morphological fractions across the four subsamples (`Lopsided', `Perturbed', `Symmetric', and `Indeterminate'). Notably, the gas morphology has different distributions in the asymmetric bar and the symmetric bar. Among the symmetric barred galaxies ($N=283$), we find 142 galaxies ($\sim50 \%$) exhibit the high gas density near the bar, and 112 galaxies ($\sim40 \%$) exhibit the hole structures. In contrast,  within the `Lopsided'-bar galaxies ($N=29$) and `Perturbed'-bar galaxies ($N=58$), the majority of galaxies ($\sim86$ and $91\%$, respectively) show the high gas density near the bar. To provide a more quantitative assessment, we calculated the central gas density within the bar radius ($\Sigma_{\text{gas, bar}}$). Figure \ref{fig:gasdensity} shows the distribution of these densities across the three subsamples. Consistent with our previous morphological findings, the gas density is systematically lower in "Symmetric" barred galaxies compared to their asymmetric counterparts ("Lopsided" and "Perturbed").

Although the sample sizes are modest, this strong contrast indicates that asymmetric bars predominantly occur in disk galaxies with abundant gas concentrated around the bar region, whereas symmetric bars are found across the full range of gas morphologies, from gas-rich to nearly gas-free systems. These findings suggest that the gas content may play a significant role in the formation of asymmetric bars. However, we note that this can be the secondary effect if the bar symmetry is affected by the other physical parameters, such as stellar mass and SFR.   

\subsection{Stellar mass \& SFR Distribution}
To determine whether stellar mass or SFR can impact the formation of asymmetric bars, we compare these parameters among the subgroups categorized by bar type. Figure \ref{fig:smass} presents the distributions of stellar masses among the four subgroups. We performed the Kolmogorov–Smirnov (K-S) test for these distributions and found that the null hypothesis that asymmetric barred galaxies (`Lopsided' and `Perturbed') share the same distributions as symmetric barred galaxies can be rejected with $p \leq 0.001$. This suggests that asymmetric bars tend to have lower stellar masses than the other classes (Fig.~\ref{fig:smass}). Interestingly, this finding aligns with the trend of more abundant gas content in asymmetric barred galaxies. \citet{semczuk_2024} demonstrated that less massive galaxies tend to have centrally concentrated gas morphology due to less efficient AGN feedback. Therefore, the high central gas density preferentially found in asymmetric barred galaxies may be the simple reflection of the stellar mass dependency of bar symmetry, which will be further discussed in \S{5.1}. 

In the distribution of SFR, we also found a discrepancy among the subgroups, as indicated by the K-S test. However, because SFR can be strongly correlated with other physical properties, such as stellar mass and gas abundance, we examined the stellar mass–SFR diagram in Figure \ref{fig:sfms} to account for this effect. As a reference line, the local star-forming main sequence (SFMS) from \citet{renzini_2015}, defined from $\sim240,000$ inactive galaxies from SDSS DR7 within $0.02<z<0.085$, which is given by   
\begin{equation}
\label{eq:SFR}
\log(\mathrm{SFR\ /\ M_\odot {\rm\ yr^{-1}}}) = (0.76 \pm 0.01)\,\log\!\left(\frac{M_\ast}{M_\odot}\right) - (7.64 \pm 0.02).
\end{equation}
Using this relation, we quantify the deviation of each galaxy from the main sequence by defining
\begin{equation}
\Delta {\rm SFR} = \log\!\left(\mathrm{SFR}\right)
        - \Bigl[\,0.76\,\log\!\left(\frac{M_\ast}{M_\odot}\right) - 7.64\,\Bigr].
\end{equation} 
This offset provides a mass-normalized measure of star formation activity relative to the local SFMS. We plot the offset to the SFMS in the right panel of Figure \ref{fig:sfms}.

Interestingly, the asymmetric barred galaxies (`Lopsided' and `Perturbed') tend to follow the SFMS within the $3\sigma$ level, whereas the other groups (`Symmetric' and `Indeterminate' barred galaxies) exhibit substantially wider distribution in the $\Delta {\rm SFR}$ (Fig.~\ref{fig:sfms}). Moreover, the `Symmetric' sample tends to have a lower $\Delta {\rm SFR}$ value compared to the other subsamples. This suggests that the asymmetric bars are more likely associated with active star formation.

\begin{figure}[t]
\centering
\includegraphics[width=0.49\textwidth]{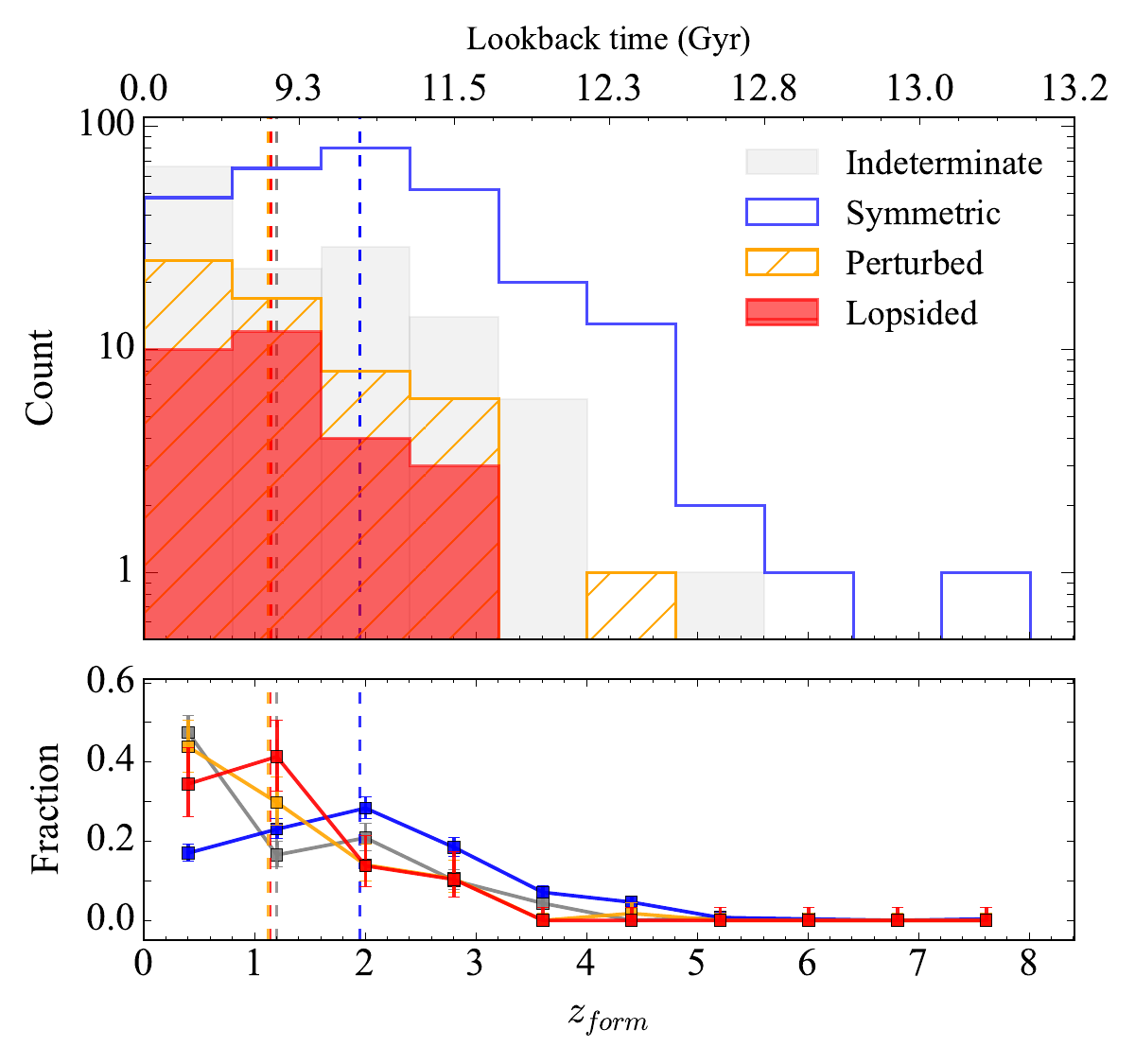}
\caption{Bar formation epochs among subgroups. Upper panel: The distribution of bar formation epochs for the four subgroups. Lower panel: The normalized fraction of objects within each epoch bin. Vertical dashed lines indicate the mean formation epoch for each corresponding subgroup.}
\vspace{0mm} 
\label{fig:bar_age}
\end{figure}

\subsection{Bar formation epoch}
To determine whether there is any physical connection between the bar age and the formation of asymmetric bars, we compared the bar formation epoch of the subgroups categorized by bar types. \citet{zana_2022} measured the $m = 2$ Fourier mode on stellar maps from the TNG50-1 simulation snapshots and provided the maximum amplitude profile ($A_{2,\max}(R)$) for the sample of barred galaxies. This time series data of $A_{2,\max}(R)$ enables us to trace the emergence of bars. Previous studies of barred galaxies based on IllustrisTNG simulations showed that $A_{2,\max}$ typically ranges from 0.2 to 0.4 \cite[][]{RosasGuevara2020, RosasGuevara2022, zana_2022}. However, different studies used different $A_{2,\max}$ threshold criteria to define barred galaxies. Specifically, \citet{RosasGuevara2020} classify disk galaxies in TNG100 with $A_{2,\max} \ge 0.3$ as strongly barred, while \citet{RosasGuevara2022} and \citet{zana_2022} adopt $A_{2,\max} \ge 0.4$ for the identification of strong bars. Motivated by these results and based on our visual inspection of the TNG50 stellar maps, we adopt the most sensitive threshold of $A_{2,\max} \ge 0.2$ to define the existence of a bar, which approximately separates weak or intermediate $m=2$ distortions from prominent bars.

To estimate the bar formation epoch for our sample, we determine the redshift (snapshot) at which the bar strength first exceeds the threshold ($A_{2,\max}(R) \ge 0.2$) within the bar in each galaxy. Figure \ref{fig:bar_age} presents the distribution of bar-formation epoch across different subgroups. Bars in the `Lopsided' and `Perturbed' tend to form more recently, with a mean formation redshift of $\langle z\rangle_{\mathrm{lop}} \simeq 1.15$ and $\langle z\rangle_{\mathrm{pert}} \simeq 1.12$, respectively. In contrast, symmetric-barred galaxies show a broader distribution with a peak around $z \approx 2.0$, and a higher mean value of $\langle z\rangle_{\mathrm{sym}} \simeq 1.95$. The bottom panel of Figure \ref{fig:bar_age} displays the fractional distribution of the four subsamples as a function of bar-formation redshift, showing that bar formation in asymmetric barred systems decreases with increasing redshift. Taken together, these trends clearly indicate that bars in asymmetric barred galaxies are systematically younger than those in symmetric barred galaxies. We note that adopting alternative thresholds for $A_{2,\rm max}$, to define the presence of a bar, does not qualitatively alter the identified trends in bar formation epochs across the subsamples.

\begin{figure}[t]
\centering
\includegraphics[width=0.49\textwidth]{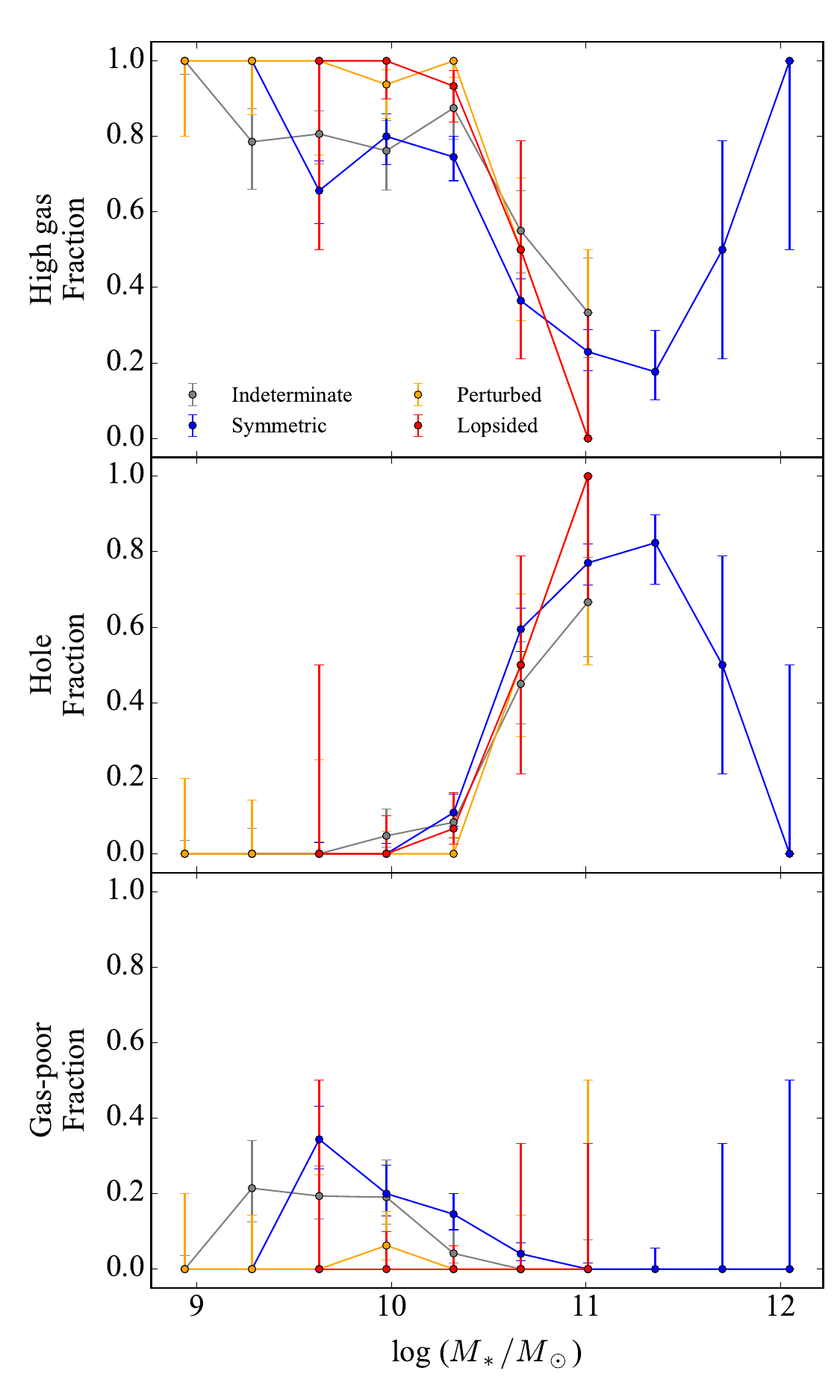}
\caption{Fractional distribution of gas morphological types as a function of stellar mass for the four subgroups. For each subgroup, we determine the relative fraction of each gas morphology within stellar mass bins. From top to bottom, panels display the fractions of high-gas, hole, and gas-poor populations.}
\vspace{0mm} 
\label{fig:gas_frac}
\end{figure}

\section{Discussion}
\label{sec:discussion}
Our visual classification reveals systematic differences in global and local properties among bar-morphology classes. In particular, galaxies hosting `Lopsided' and `Perturbed' bars mostly show enhanced gas surface density around the bar region, tend to lie close to the local SFMS, and exhibit more recent bar formation epochs than systems with `Symmetric' bars. Below, we discuss physical scenarios that can produce or sustain asymmetric barred galaxies and propose diagnostics based on observation and simulation.

\begin{figure}[t]
\centering
\includegraphics[width=0.49\textwidth]{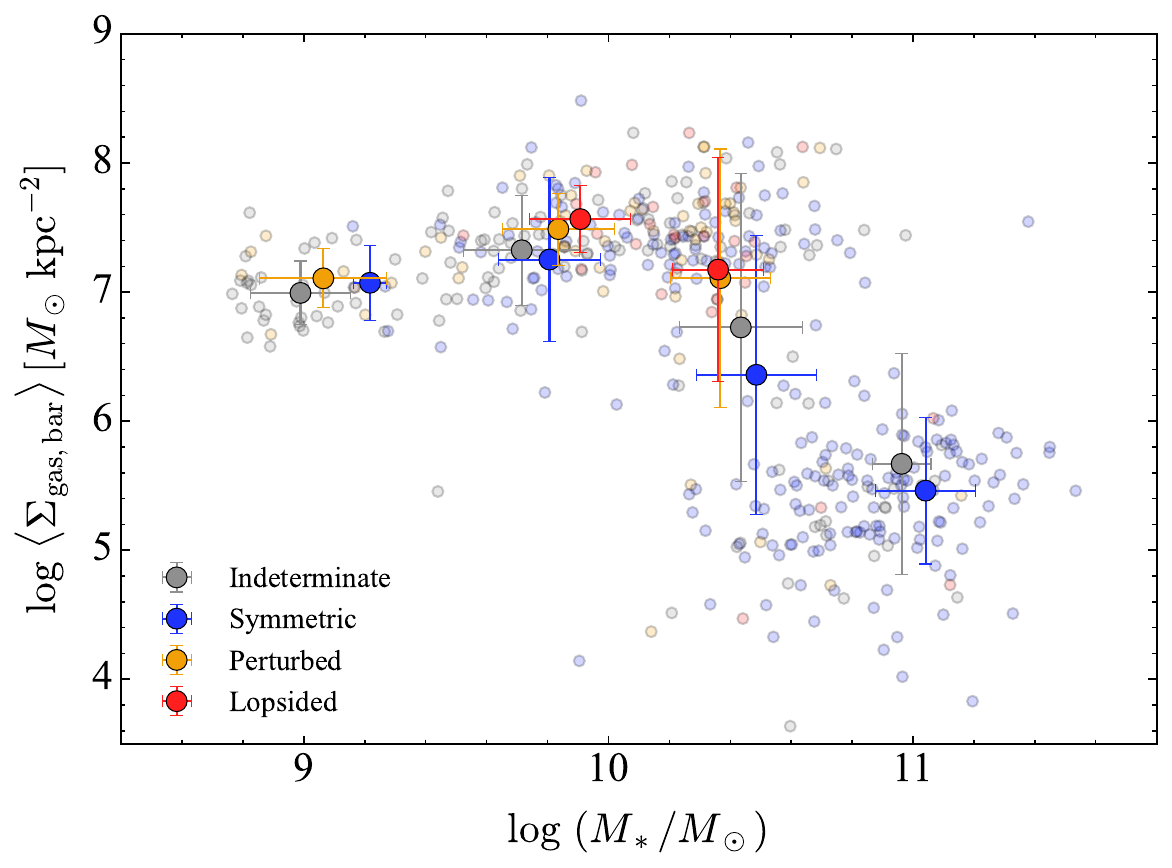}
\caption{Central gas density as a function of stellar mass for the four subgroups. Filled circles with error bars indicate binned values and their associated dispersions.}
\vspace{0mm} 
\label{fig:gas_frac2}
\end{figure}

\subsection{Gas-driven Scenario}
\label{sec:gasdriven}
Our results indicate that asymmetric bars tend to reside in systems with high gas density around the bar, whereas symmetric bars span a broader range of gas morphologies (including galaxies with a gas hole and gas-poor disks). 
This finding may suggest that abundant gas in the bar vicinity is not merely incidental but may be relevant to the formation of asymmetric structures. Cosmological gas inflow can promote enhanced star formation, leading to the formation of massive gas clumps and localized overdensities and thus asymmetric disks \cite[][]{bournaud_2005, bournaud_2007, inoue_2012}. These structures can gravitationally interact with the stellar component, perturb the bar potential \citep{ceverino_2010}, and thus potentially give rise to bar asymmetries. Thus, having abundant gas in the bar region may be crucial for maintaining or amplifying bar lopsidedness. 

\begin{figure*}[t]
\centering
\includegraphics[width=1\textwidth]{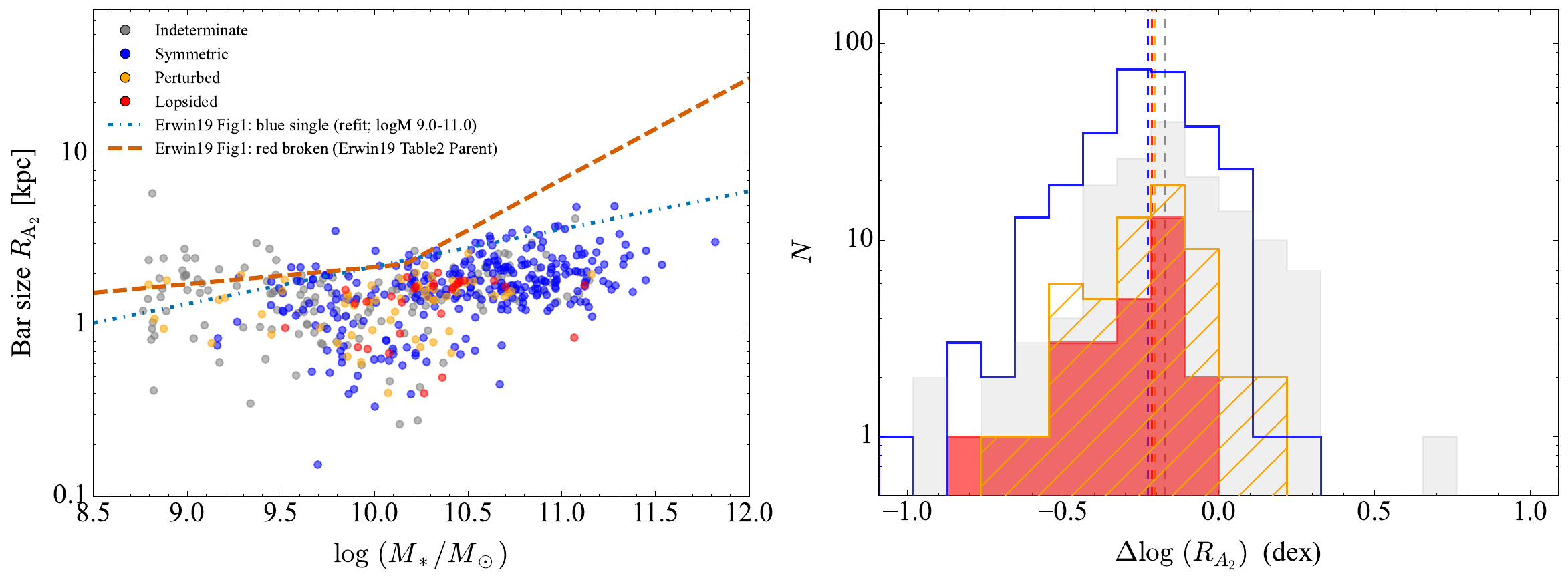}
\caption{Comparison between the bar size and stellar mass. 
Left: Distributions of bar size as a function of stellar mass. The red-dashed and blue-dotted lines represent the broken power-law and single power-law relations, respectively, for local barred galaxies from \citet{erwin_2019}. The four subgroups are color-coded as indicated in the legend. Right: Distribution of the bar size offset from the local single power-law relation. Vertical dashed lines indicate the mean offset for each subgroup.}
\vspace{0mm} 
\label{fig:bar_sm}
\end{figure*}

Figure \ref{fig:gas_frac} presents the fraction of gas morphologies for the four subgroups in each stellar mass bin, accounting for the dependence of gas properties on stellar mass. The overall trend of the gas-morphology fractions as a function of stellar mass is broadly similar among the four bar-morphology classes: systems tend to be classified as high-gas at lower $M_{\ast}$, while the hole- and/or gas-poor fractions increase toward higher $M_\ast$. The fact that this mass-dependent behavior appears in all four groups indicates that the gas morphology in our classification primarily traces global galaxy properties tied to stellar mass \cite[e.g.,][]{semczuk_2024}, rather than uniquely tracking the presence of an asymmetric bar. Substituting gas morphology with a quantitative measurement of the gas surface density within the bar ($\Sigma_{\text{gas, bar}}$) confirms that stellar mass remains the primary parameter governing the observed trends (Fig.~\ref{fig:gas_frac2}).
Taken together, our results suggest a picture in which stellar mass is the primary parameter associated with bar lopsidedness, with gas content acting as a correlated component rather than the dominant physical origin. 

However, this conclusion does not sufficiently explain our finding that galaxies with asymmetric bars tend to follow the SFMS, whereas those with symmetric bars lie below it. This discrepancy suggests that star formation may be further enhanced in asymmetric bars, even after accounting for gas abundance. While the physical origin of this enhancement remains unclear, it is likely associated with the external triggers of bar asymmetry, such as tidal interactions with neighboring galaxies (see Section~\ref{sec:ageeffect}).

\subsection{Age Effect}
\label{sec:ageeffect}
Our analysis of the bar-formation epoch suggests that asymmetric bars are systematically younger (i.e., more recently formed) than symmetric bars. A natural interpretation is that lopsidedness may represent an early evolutionary phase, occurring during or shortly after the bar formation. In this picture, a recently formed bar has not yet fully settled to have a symmetric orbital structure: small pre-existing asymmetries in the disk (e.g., mild m = 1 distortions) and stochastic clumps which are frequently observed in high-z \cite[e.g.,][]{conselice_2014, elmergreen_2005, foster_2011, kalita_2025} may affect the morphological shapes of the bar.

As a galaxy evolves, a bar experiences buckling instability 
\cite[e.g.,][]{raha_1991, martinez_2004, athanassoula_2005, debattista_2006}. \citet{lokas_2021} find that the asymmetry of the bar disappears after the buckling. This also indicates that asymmetric bars are in their early evolutionary phase. This evolutionary view is also consistent with our result that asymmetric bars tend to lie close to the local SFMS from \citet{renzini_2015}, suggesting that they are hosted by actively star-forming disks which have ample gas.

However, the fraction of asymmetric bars in this study is significantly higher than that observed in local galaxies \cite[e.g.,][]{vandermarel_2001, jacyszyn-Dobrzeniecka_2016, patra_2019, cuomo_2022}. This discrepancy suggests that bar asymmetry in the TNG50 simulation may be amplified by external mechanisms, such as tidal interactions with neighboring galaxies. Consequently, bar evolution within TNG50 may differ from bar formation processes in the real Universe, potentially biasing our results toward specific environmental triggers and limiting the broader interpretation of these findings (see Section~\ref{sec:barlength} for further discussion).

\subsection{Bar Length}
\label{sec:barlength}
We found that asymmetric barred galaxies are likely to exhibit younger bars and are preferentially hosted by less massive galaxies than symmetric barred galaxies. Previous studies suggest that bar length increases over time \cite[e.g.,][]{athanassoula_2013} and depends on the stellar mass of the host galaxy \cite[e.g.,][]{diaz-garcia_2016, erwin_2019, kim_2021}. Combining these results with our findings, we suggest that bar length may vary significantly across the four subgroups. To assess this hypothesis, we examine the bar sizes in our sample while accounting for their dependence on stellar mass. We adopt $R_{A_2}$ as a proxy of the bar size. Contrary to our initial prediction, bar size does not vary significantly across the four subgroups. This suggests that bar size is not a primary driver of the observed bar lopsidedness.
We also found that bars in TNG50-1 are systematically smaller than those in the barred galaxies in the local Universe at a given stellar mass (Fig.~\ref{fig:bar_sm}; \citealp{erwin_2019}). While investigating the physical origin of this discrepancy is beyond the scope of this study, it may arise from either simulation limitations in capturing bar formation and evolution or systematic differences in bar length measurement methodologies. 

\begin{figure}[t]
\centering
\includegraphics[width=0.49\textwidth]{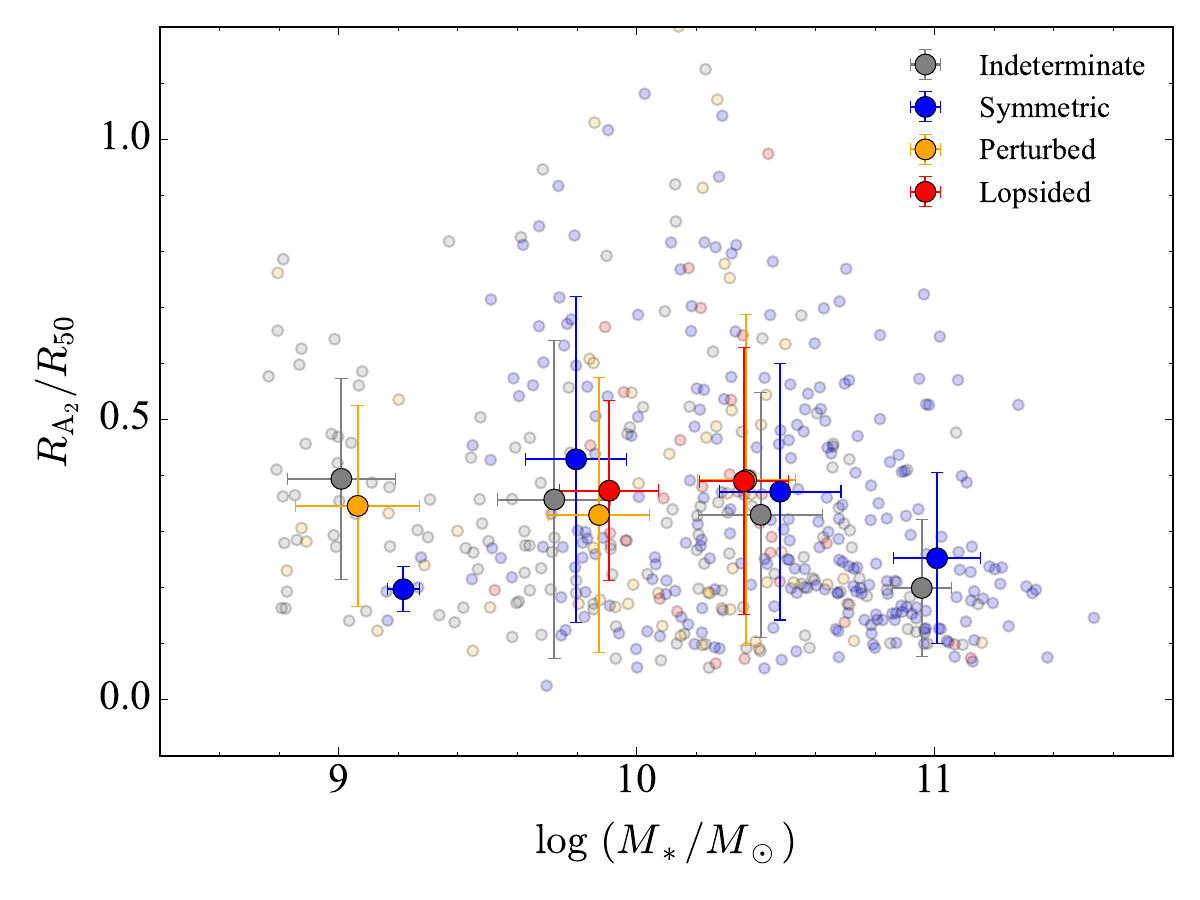}
\caption{Ratio of bar size to the stellar half-mass radius as a function of stellar mass. Filled circles with error bars indicate binned values and their associated dispersions.}
\vspace{0mm} 
\label{fig:bar_hmr}
\end{figure}

The bar length is also known to correlate with galaxy size \cite[e.g.,][]{erwin_2019}, which should be taken into account for reliable assessments. Here, we used the stellar half-mass radius ($R_{50}$) as an indicator of galaxy size. We examined the relative bar length to $R_{50}$ as a function of the stellar mass and again found no significant difference among the four subgroups (Fig.~\ref{fig:bar_hmr}), supporting the result from the absolute bar length. In conclusion, when accounting for stellar mass, the bar length shows no physical correlation with bar asymmetry.

\section{Conclusion} \label{sec:conclusion}
To investigate the physical origin of asymmetric bars, we utilized 519 barred galaxies from the cosmological simulation, TNG50-1. We visually classified the sample into subgroups based on the degree of asymmetry using the stellar mass maps. Bars associated with asymmetric 2-D profiles are categorized as `Lopsided' if the asymmetry is prominent and extended to the outer part of the bar and `Perturbed' if the asymmetry is moderate and localized within the inner part of the bar. As a counterpart of asymmetric bars, samples without any clear signature of asymmetry in the bar profile are categorized as `Symmetric'. Finally, ambiguous cases primarily attributed to the weak surface density within bars are `Indeterminate'. By examining the physical properties of four subgroups, we attempted to investigate the origin of the asymmetric bars in the simulation data and reached the following conclusions:

\begin{itemize}

\item Of the barred galaxies identified in the TNG50-1 simulation, 5.6\%, 11.2\%, 54.5\%, and 28.7\% are classified as `Lopsided', `Perturbed', `Symmetric', and `Indeterminate', respectively.

\item Asymmetric bars (categorized as `Lopsided' or `Perturbed') are preferentially hosted by lower-mass galaxies compared to their symmetric counterparts. Furthermore, a vast majority ($\approx 90\%$) of asymmetric bars exhibit enhanced gas densities in the bar region, whereas only half of symmetric bars show similar gas distributions. Approximately 40$\%$ of symmetric bars are associated with central gas cavities (holes). We find, however, that the dependence of gas morphology on bar asymmetry largely mirrors the underlying stellar mass–gas morphology relation. This suggests that stellar mass, rather than local gas dynamics, is the primary driver of bar asymmetry.

\item In the $M_{\ast}$-SFR plane, symmetric barred galaxies are predominantly located along the star-forming main sequence (SFMS), whereas symmetric barred galaxies typically reside below it. This distribution indicates a physical connection between bar asymmetry and active star formation compared to their symmetric counterparts.
 
\item Analysis of the $m=2$ Fourier mode amplitudes reveals that bars in asymmetric galaxies are systematically younger than their symmetric counterparts. This suggests a secular evolutionary link, in which bar symmetry is physically coupled to the aging process, with asymmetric bars representing a transient phase that eventually relaxes into a symmetric bar. 

\item Absolute and relative bar sizes as a function of stellar mass show no significant difference between asymmetric and symmetric bars, suggesting no physical connection between bar size and symmetry. 

\item While our analysis provides insight into the physical origin of asymmetric bars, the fraction identified in our sample ($\sim17\%$) is significantly higher than the observed rarity of such features in the low-$z$ Universe. This discrepancy suggests that asymmetric features in TNG50-1 may be enhanced by external mechanisms, such as tidal interactions with neighboring galaxies. Consequently, these results should be interpreted with caution.

\end{itemize}


\acknowledgments

We are grateful to the anonymous referee for constructive suggestions that greatly improve the quality of the manuscript. This work was supported by the National Research Foundation of Korea (NRF) grant funded by the Korean government (MSIT) (Nos. RS-2024-00347548 and RS-2025-16066624) and the Yonsei University Research Fund of 2025-22-1785. TK was supported by the National Research Foundation of Korea (NRF) grant funded by the Korean government (MSIT) (No. RS-2025-25399934).

\bibliography{bar}{}

\begin{thebibliography}{}
\expandafter\ifx\csname natexlab\endcsname\relax\def\natexlab#1{#1}\fi
\providecommand{\url}[1]{\href{#1}{#1}}
\providecommand{\dodoi}[1]{doi:~\href{http://doi.org/#1}{\nolinkurl{#1}}}
\providecommand{\doeprint}[1]{\href{http://ascl.net/#1}{\nolinkurl{http://ascl.net/#1}}}
\providecommand{\doarXiv}[1]{\href{https://arxiv.org/abs/#1}{\nolinkurl{https://arxiv.org/abs/#1}}}
\providecommand{\dodoilink}[2]{\href{http://doi.org/#1}{#2}}
\providecommand{\doadslink}[2]{\href{#1}{#2}}

\bibitem[{{Aguerri} {et~al.}(2009){Aguerri}, {M{\'e}ndez-Abreu}, \& {Corsini}}]{aguerri_2009}
{Aguerri}, J.~A.~L., {M{\'e}ndez-Abreu}, J., \& {Corsini}, E.~M. 2009, \aap, 495, 491

\bibitem[{{Alonso} {et~al.}(2018){Alonso}, {Coldwell}, {Duplancic}, {Mesa}, \& {Lambas}}]{alonso_2018}
{Alonso}, S., {Coldwell}, G., {Duplancic}, F., {Mesa}, V., \& {Lambas}, D.~G. 2018, \aap, 618, A149

\bibitem[{{Athanassoula}(1992)}]{athanassoula_1992a}
{Athanassoula}, E. 1992, \mnras, 259, 328

\bibitem[{{Athanassoula}(1996)}]{athanassoula_1996}
{Athanassoula}, E. 1996, in Astronomical Society of the Pacific Conference Series, Vol.~91, IAU Colloquium 157: Barred Galaxies, ed. R.~{Buta}, D.~A. {Crocker}, \& B.~G. {Elmegreen}, 309

\bibitem[{{Athanassoula}(2005)}]{athanassoula_2005}
{Athanassoula}, E. 2005, \mnras, 358, 1477

\bibitem[{{Athanassoula} {et~al.}(2013){Athanassoula}, {Machado}, \& {Rodionov}}]{athanassoula_2013}
{Athanassoula}, E., {Machado}, R. E.~G., \& {Rodionov}, S.~A. 2013, \mnras, 429, 1949

\bibitem[{{Athanassoula} {et~al.}(2009){Athanassoula}, {Romero-G{\'o}mez}, {Bosma}, \& {Masdemont}}]{athanassoula_2009}
{Athanassoula}, E., {Romero-G{\'o}mez}, M., {Bosma}, A., \& {Masdemont}, J.~J. 2009, \mnras, 400, 1706

\bibitem[{{Baes} {et~al.}(2011){Baes}, {Verstappen}, {De Looze}, {Fritz}, {Saftly}, {Vidal P{\'e}rez}, {Stalevski}, \& {Valcke}}]{baes_2011}
{Baes}, M., {Verstappen}, J., {De Looze}, I., {et~al.} 2011, \apjs, 196, 22

\bibitem[{{Bekki}(2009)}]{bekki_2009}
{Bekki}, K. 2009, \mnras, 393, L60

\bibitem[{{Berentzen} {et~al.}(2003){Berentzen}, {Athanassoula}, {Heller}, \& {Fricke}}]{berentzen_2003}
{Berentzen}, I., {Athanassoula}, E., {Heller}, C.~H., \& {Fricke}, K.~J. 2003, \mnras, 341, 343

\bibitem[{{Berentzen} {et~al.}(2007){Berentzen}, {Shlosman}, {Martinez-Valpuesta}, \& {Heller}}]{berentzen_2007}
{Berentzen}, I., {Shlosman}, I., {Martinez-Valpuesta}, I., \& {Heller}, C.~H. 2007, \apj, 666, 189

\bibitem[{{Bournaud} {et~al.}(2005){Bournaud}, {Combes}, \& {Semelin}}]{bournaud_2005}
{Bournaud}, F., {Combes}, F., \& {Semelin}, B. 2005, \mnras, 364, L18

\bibitem[{Bournaud {et~al.}(2007)Bournaud, Elmegreen, \& Elmegreen}]{bournaud_2007}
Bournaud, F., Elmegreen, B.~G., \& Elmegreen, D.~M. 2007, The Astrophysical Journal, 670, 237

\bibitem[{{Buta} {et~al.}(2005){Buta}, {Vasylyev}, {Salo}, \& {Laurikainen}}]{buta_2005}
{Buta}, R., {Vasylyev}, S., {Salo}, H., \& {Laurikainen}, E. 2005, \aj, 130, 506

\bibitem[{{Buta}(2017)}]{buta_2017b}
{Buta}, R.~J. 2017, \mnras, 470, 3819

\bibitem[{{Buta} {et~al.}(2009){Buta}, {Knapen}, {Elmegreen}, {Salo}, {Laurikainen}, {Elmegreen}, {Puerari}, \& {Block}}]{buta_2009a}
{Buta}, R.~J., {Knapen}, J.~H., {Elmegreen}, B.~G., {et~al.} 2009, \aj, 137, 4487

\bibitem[{{Buta} {et~al.}(2015){Buta}, {Sheth}, {Athanassoula}, {Bosma}, {Knapen}, {Laurikainen}, {Salo}, {Elmegreen}, {Ho}, {Zaritsky}, {Courtois}, {Hinz}, {Mu{\~n}oz-Mateos}, {Kim}, {Regan}, {Gadotti}, {Gil de Paz}, {Laine}, {Men{\'e}ndez-Delmestre}, {Comer{\'o}n}, {Erroz Ferrer}, {Seibert}, {Mizusawa}, {Holwerda}, \& {Madore}}]{buta_2015}
{Buta}, R.~J., {Sheth}, K., {Athanassoula}, E., {et~al.} 2015, \apjs, 217, 32

\bibitem[{{Camps} \& {Baes}(2015)}]{camps_2015}
{Camps}, P., \& {Baes}, M. 2015, Astronomy and Computing, 9, 20

\bibitem[{{Ceverino} {et~al.}(2010){Ceverino}, {Dekel}, \& {Bournaud}}]{ceverino_2010}
{Ceverino}, D., {Dekel}, A., \& {Bournaud}, F. 2010, \mnras, 404, 2151

\bibitem[{{Cheung} {et~al.}(2013){Cheung}, {Athanassoula}, {Masters}, {Nichol}, {Bosma}, {Bell}, {Faber}, {Koo}, {Lintott}, {Melvin}, {Schawinski}, {Skibba}, \& {Willett}}]{cheung_2013}
{Cheung}, E., {Athanassoula}, E., {Masters}, K.~L., {et~al.} 2013, \apj, 779, 162

\bibitem[{{Cheung} {et~al.}(2015){Cheung}, {Trump}, {Athanassoula}, {Bamford}, {Bell}, {Bosma}, {Cardamone}, {Casteels}, {Faber}, {Fang}, {Fortson}, {Kocevski}, {Koo}, {Laine}, {Lintott}, {Masters}, {Melvin}, {Nichol}, {Schawinski}, {Simmons}, {Smethurst}, \& {Willett}}]{cheung_2015}
{Cheung}, E., {Trump}, J.~R., {Athanassoula}, E., {et~al.} 2015, \mnras, 447, 506

\bibitem[{{Cisternas} {et~al.}(2015){Cisternas}, {Sheth}, {Salvato}, {Knapen}, {Civano}, \& {Santini}}]{cisternas_2015}
{Cisternas}, M., {Sheth}, K., {Salvato}, M., {et~al.} 2015, \apj, 802, 137

\bibitem[{{Col{\'{\i}}n} \& {Athanassoula}(1989)}]{colin_1989}
{Col{\'{\i}}n}, J., \& {Athanassoula}, E. 1989, Astronomy and Astrophysics, 214, 99

\bibitem[{{Conselice}(2014)}]{conselice_2014}
{Conselice}, C.~J. 2014, \araa, 52, 291

\bibitem[{{Cuomo} {et~al.}(2022){Cuomo}, {Corsini}, {Morelli}, {Aguerri}, {Lee}, {Coccato}, {Pizzella}, {Buttitta}, \& {Gasparri}}]{cuomo_2022}
{Cuomo}, V., {Corsini}, E.~M., {Morelli}, L., {et~al.} 2022, \mnras, 516, L24

\bibitem[{{de Lorenzo-C{\'a}ceres} {et~al.}(2019){de Lorenzo-C{\'a}ceres}, {S{\'a}nchez-Bl{\'a}zquez}, {M{\'e}ndez-Abreu}, {Gadotti}, {Falc{\'o}n-Barroso}, {Mart{\'\i}nez-Valpuesta}, {Coelho}, {Fragkoudi}, {Husemann}, {Leaman}, {P{\'e}rez}, {Querejeta}, {Seidel}, \& {van de Ven}}]{delorenzo_2019}
{de Lorenzo-C{\'a}ceres}, A., {S{\'a}nchez-Bl{\'a}zquez}, P., {M{\'e}ndez-Abreu}, J., {et~al.} 2019, \mnras, 484, 5296

\bibitem[{{de S{\'a}-Freitas} {et~al.}(2023){de S{\'a}-Freitas}, {Fragkoudi}, {Gadotti}, {Falc{\'o}n-Barroso}, {Bittner}, {S{\'a}nchez-Bl{\'a}zquez}, {van de Ven}, {Bieri}, {Coccato}, {Coelho}, {Fahrion}, {Gon{\c{c}}alves}, {Kim}, {de Lorenzo-C{\'a}ceres}, {Martig}, {Mart{\'\i}n-Navarro}, {Mendez-Abreu}, {Neumann}, \& {Querejeta}}]{desafreitas_2023}
{de S{\'a}-Freitas}, C., {Fragkoudi}, F., {Gadotti}, D.~A., {et~al.} 2023, \aap, 671, A8

\bibitem[{{de Swardt} {et~al.}(2015){de Swardt}, {Sheth}, {Kim}, {Pardy}, {D'Onghia}, {Wilcots}, {Hinz}, {Mu{\~n}oz-Mateos}, {Regan}, {Athanassoula}, {Bosma}, {Buta}, {Cisternas}, {Comer{\'o}n}, {Gadotti}, {Gil de Paz}, {Jarrett}, {Elmegreen}, {Erroz-Ferrer}, {Ho}, {Knapen}, {Laine}, {Laurikainen}, {Madore}, {Meidt}, {Men{\'e}ndez-Delmestre}, {Peng}, {Salo}, {Schinnerer}, \& {Zaritsky}}]{deswardt_2015}
{de Swardt}, B., {Sheth}, K., {Kim}, T., {et~al.} 2015, \apj, 808, 90

\bibitem[{{Debattista} {et~al.}(2006){Debattista}, {Mayer}, {Carollo}, {Moore}, {Wadsley}, \& {Quinn}}]{debattista_2006}
{Debattista}, V.~P., {Mayer}, L., {Carollo}, C.~M., {et~al.} 2006, \apj, 645, 209

\bibitem[{{D{\'\i}az-Garc{\'\i}a} {et~al.}(2016){D{\'\i}az-Garc{\'\i}a}, {Salo}, {Laurikainen}, \& {Herrera-Endoqui}}]{diaz-garcia_2016}
{D{\'\i}az-Garc{\'\i}a}, S., {Salo}, H., {Laurikainen}, E., \& {Herrera-Endoqui}, M. 2016, \aap, 587, A160

\bibitem[{{Ellison} {et~al.}(2011){Ellison}, {Nair}, {Patton}, {Scudder}, {Mendel}, \& {Simard}}]{ellison_2011}
{Ellison}, S.~L., {Nair}, P., {Patton}, D.~R., {et~al.} 2011, \mnras, 416, 2182

\bibitem[{{Elmegreen} \& {Elmegreen}(2005)}]{elmergreen_2005}
{Elmegreen}, B.~G., \& {Elmegreen}, D.~M. 2005, \apj, 627, 632

\bibitem[{{Erwin}(2019)}]{erwin_2019}
{Erwin}, P. 2019, \mnras, 489, 3553

\bibitem[{{Eskridge} {et~al.}(2000){Eskridge}, {Frogel}, {Pogge}, {Quillen}, {Davies}, {DePoy}, {Houdashelt}, {Kuchinski}, {Ram{\'\i}rez}, {Sellgren}, {Terndrup}, \& {Tiede}}]{eskridge_2000}
{Eskridge}, P.~B., {Frogel}, J.~A., {Pogge}, R.~W., {et~al.} 2000, \aj, 119, 536

\bibitem[{{F{\"o}rster Schreiber} {et~al.}(2011){F{\"o}rster Schreiber}, {Shapley}, {Genzel}, {Bouch{\'e}}, {Cresci}, {Davies}, {Erb}, {Genel}, {Lutz}, {Newman}, {Shapiro}, {Steidel}, {Sternberg}, \& {Tacconi}}]{foster_2011}
{F{\"o}rster Schreiber}, N.~M., {Shapley}, A.~E., {Genzel}, R., {et~al.} 2011, \apj, 739, 45

\bibitem[{{Fragkoudi} {et~al.}(2016){Fragkoudi}, {Athanassoula}, \& {Bosma}}]{fragkoudi_2016}
{Fragkoudi}, F., {Athanassoula}, E., \& {Bosma}, A. 2016, \mnras, 462, L41

\bibitem[{{Gadotti}(2009)}]{gadotti_2009}
{Gadotti}, D.~A. 2009, \mnras, 393, 1531

\bibitem[{{Gadotti} {et~al.}(2020){Gadotti}, {Bittner}, {Falc{\'o}n-Barroso}, {M{\'e}ndez-Abreu}, {Kim}, {Fragkoudi}, {de Lorenzo-C{\'a}ceres}, {Leaman}, {Neumann}, {Querejeta}, {S{\'a}nchez-Bl{\'a}zquez}, {Martig}, {Mart{\'\i}n-Navarro}, {P{\'e}rez}, {Seidel}, \& {van de Ven}}]{gadotti_2020}
{Gadotti}, D.~A., {Bittner}, A., {Falc{\'o}n-Barroso}, J., {et~al.} 2020, \aap, 643, A14

\bibitem[{{Galloway} {et~al.}(2015){Galloway}, {Willett}, {Fortson}, {Cardamone}, {Schawinski}, {Cheung}, {Lintott}, {Masters}, {Melvin}, \& {Simmons}}]{galloway_2015}
{Galloway}, M.~A., {Willett}, K.~W., {Fortson}, L.~F., {et~al.} 2015, \mnras, 448, 3442

\bibitem[{{Ghosh} {et~al.}(2024){Ghosh}, {Gadotti}, {Fragkoudi}, {Nagpal}, {Di Matteo}, \& {Cuomo}}]{ghosh_2024b}
{Ghosh}, S., {Gadotti}, D.~A., {Fragkoudi}, F., {et~al.} 2024, \mnras, 532, 4570

\bibitem[{{Ho} {et~al.}(1997){Ho}, {Filippenko}, \& {Sargent}}]{ho_1997}
{Ho}, L.~C., {Filippenko}, A.~V., \& {Sargent}, W. L.~W. 1997, \apjs, 112, 315

\bibitem[{{Hunt} \& {Malkan}(1999)}]{hunt_1999}
{Hunt}, L.~K., \& {Malkan}, M.~A. 1999, \apj, 516, 660

\bibitem[{{Inoue} \& {Saitoh}(2012)}]{inoue_2012}
{Inoue}, S., \& {Saitoh}, T.~R. 2012, \mnras, 422, 1902

\bibitem[{{Jacyszyn-Dobrzeniecka} {et~al.}(2016){Jacyszyn-Dobrzeniecka}, {Skowron}, {Mr{\'o}z}, {Skowron}, {Soszy{\'n}ski}, {Udalski}, {Pietrukowicz}, {Koz{\l}owski}, {Wyrzykowski}, {Poleski}, {Pawlak}, {Szyma{\'n}ski}, \& {Ulaczyk}}]{jacyszyn-Dobrzeniecka_2016}
{Jacyszyn-Dobrzeniecka}, A.~M., {Skowron}, D.~M., {Mr{\'o}z}, P., {et~al.} 2016, \actaa, 66, 149

\bibitem[{{James} \& {Percival}(2018)}]{james_2018}
{James}, P.~A., \& {Percival}, S.~M. 2018, \mnras, 474, 3101

\bibitem[{{Jogee} {et~al.}(2005){Jogee}, {Scoville}, \& {Kenney}}]{jogee_2005}
{Jogee}, S., {Scoville}, N., \& {Kenney}, J. D.~P. 2005, \apj, 630, 837

\bibitem[{{Kalita} {et~al.}(2025){Kalita}, {Silverman}, {Daddi}, {Mercier}, {Ho}, \& {Ding}}]{kalita_2025}
{Kalita}, B.~S., {Silverman}, J.~D., {Daddi}, E., {et~al.} 2025, \mnras, 537, 402

\bibitem[{{Kim} {et~al.}(2021){Kim}, {Athanassoula}, {Sheth}, {Bosma}, {Park}, {Lee}, \& {Ann}}]{kim_2021}
{Kim}, T., {Athanassoula}, E., {Sheth}, K., {et~al.} 2021, \apj, 922, 196

\bibitem[{{Kim} {et~al.}(2025){Kim}, {Gadotti}, {Park}, {Lee}, {Fragkoudi}, {Kim}, \& {Kim}}]{kim_2025b}
{Kim}, T., {Gadotti}, D.~A., {Park}, M.-g., {et~al.} 2025, \apj, 994, 105

\bibitem[{{Kim} {et~al.}(2012){Kim}, {Seo}, \& {Kim}}]{kim_2012}
{Kim}, W.-T., {Seo}, W.-Y., \& {Kim}, Y. 2012, \apj, 758, 14

\bibitem[{{Kormendy} \& {Fisher}(2005)}]{kormendy_2005}
{Kormendy}, J., \& {Fisher}, D.~B. 2005, in Revista Mexicana de Astronomia y Astrofisica Conference Series, Vol.~23, Revista Mexicana de Astronomia y Astrofisica Conference Series, ed. S.~{Torres-Peimbert} \& G.~{MacAlpine}, 101--108, \dodoi{10.48550/arXiv.astro-ph/0507525}

\bibitem[{{Kormendy} \& {Kennicutt}(2004)}]{kormendy_2004}
{Kormendy}, J., \& {Kennicutt}, Jr., R.~C. 2004, \araa, 42, 603

\bibitem[{{Kruk} {et~al.}(2017){Kruk}, {Lintott}, {Simmons}, {Bamford}, {Cardamone}, {Fortson}, {Hart}, {H{\"a}u{\ss}ler}, {Masters}, {Nichol}, {Schawinski}, \& {Smethurst}}]{kruk_2017}
{Kruk}, S.~J., {Lintott}, C.~J., {Simmons}, B.~D., {et~al.} 2017, \mnras, 469, 3363

\bibitem[{{Kuno} {et~al.}(2007){Kuno}, {Sato}, {Nakanishi}, {Hirota}, {Tosaki}, {Shioya}, {Sorai}, {Nakai}, {Nishiyama}, \& {Vila-Vilar{\'o}}}]{kuno_2007}
{Kuno}, N., {Sato}, N., {Nakanishi}, H., {et~al.} 2007, \pasj, 59, 117

\bibitem[{{Laurikainen} \& {Salo}(2002)}]{laurikainen_2002}
{Laurikainen}, E., \& {Salo}, H. 2002, \mnras, 337, 1118

\bibitem[{{Lee} {et~al.}(2012){Lee}, {Woo}, {Lee}, {Hwang}, {Lee}, {Sohn}, \& {Lee}}]{lee_2012}
{Lee}, G.-H., {Woo}, J.-H., {Lee}, M.~G., {et~al.} 2012, \apj, 750, 141

\bibitem[{{Lee} {et~al.}(2020){Lee}, {Park}, {Ann}, {Kim}, \& {Seo}}]{lee_2020}
{Lee}, Y.~H., {Park}, M.-G., {Ann}, H.~B., {Kim}, T., \& {Seo}, W.-Y. 2020, \apj, 899, 84

\bibitem[{{{\L}okas}(2021)}]{lokas_2021}
{{\L}okas}, E.~L. 2021, \aap, 655, A97

\bibitem[{{Martinez-Valpuesta} \& {Shlosman}(2004)}]{martinez_2004}
{Martinez-Valpuesta}, I., \& {Shlosman}, I. 2004, \apjl, 613, L29

\bibitem[{{Masters} {et~al.}(2011){Masters}, {Nichol}, {Hoyle}, {Lintott}, {Bamford}, {Edmondson}, {Fortson}, {Keel}, {Schawinski}, {Smith}, \& {Thomas}}]{masters_2011}
{Masters}, K.~L., {Nichol}, R.~C., {Hoyle}, B., {et~al.} 2011, \mnras, 411, 2026

\bibitem[{{Men{\'e}ndez-Delmestre} {et~al.}(2007){Men{\'e}ndez-Delmestre}, {Sheth}, {Schinnerer}, {Jarrett}, \& {Scoville}}]{menedez_2007}
{Men{\'e}ndez-Delmestre}, K., {Sheth}, K., {Schinnerer}, E., {Jarrett}, T.~H., \& {Scoville}, N.~Z. 2007, \apj, 657, 790

\bibitem[{{Nelson} {et~al.}(2019){Nelson}, {Pillepich}, {Springel}, {Pakmor}, {Weinberger}, {Genel}, {Torrey}, {Vogelsberger}, {Marinacci}, \& {Hernquist}}]{nelson_2019}
{Nelson}, D., {Pillepich}, A., {Springel}, V., {et~al.} 2019, \mnras, 490, 3234

\bibitem[{{Newnham} {et~al.}(2020){Newnham}, {Hess}, {Masters}, {Kruk}, {Penny}, {Lingard}, \& {Smethurst}}]{newnham_2020}
{Newnham}, L., {Hess}, K.~M., {Masters}, K.~L., {et~al.} 2020, \mnras, 492, 4697

\bibitem[{{Odewahn}(1996)}]{odewahn_1996}
{Odewahn}, S.~C. 1996, Astronomical Society of the Pacific Conference Series, 106, 139

\bibitem[{{Oh} {et~al.}(2012){Oh}, {Oh}, \& {Yi}}]{oh_2012}
{Oh}, S., {Oh}, K., \& {Yi}, S.~K. 2012, \apjs, 198, 4

\bibitem[{{Ohta} {et~al.}(1990){Ohta}, {Hamabe}, \& {Wakamatsu}}]{ohta_1990}
{Ohta}, K., {Hamabe}, M., \& {Wakamatsu}, K.-I. 1990, \apj, 357, 71

\bibitem[{{Pardy} {et~al.}(2016){Pardy}, {D'Onghia}, {Athanassoula}, {Wilcots}, \& {Sheth}}]{pardy_2016}
{Pardy}, S.~A., {D'Onghia}, E., {Athanassoula}, E., {Wilcots}, E.~M., \& {Sheth}, K. 2016, \apj, 827, 149

\bibitem[{{Patra} \& {Jog}(2019)}]{patra_2019}
{Patra}, N.~N., \& {Jog}, C.~J. 2019, \mnras, 488, 4942

\bibitem[{{Peschken} \& {{\L}okas}(2019)}]{Peschken2019}
{Peschken}, N., \& {{\L}okas}, E.~L. 2019, \mnras, 483, 2721

\bibitem[{{Pillepich} {et~al.}(2018){Pillepich}, {Springel}, {Nelson}, {Genel}, {Naiman}, {Pakmor}, {Hernquist}, {Torrey}, {Vogelsberger}, {Weinberger}, \& {Marinacci}}]{Pillepich2018}
{Pillepich}, A., {Springel}, V., {Nelson}, D., {et~al.} 2018, \mnras, 473, 4077

\bibitem[{{Pillepich} {et~al.}(2019){Pillepich}, {Nelson}, {Springel}, {Pakmor}, {Torrey}, {Weinberger}, {Vogelsberger}, {Marinacci}, {Genel}, {van der Wel}, \& {Hernquist}}]{pillepich_2019}
{Pillepich}, A., {Nelson}, D., {Springel}, V., {et~al.} 2019, \mnras, 490, 3196

\bibitem[{{Planck Collaboration} {et~al.}(2016){Planck Collaboration}, {Ade}, {Aghanim}, {Arnaud}, {Ashdown}, {Aumont}, {Baccigalupi}, {Banday}, {Barreiro}, {Bartlett}, {Bartolo}, {Battaner}, {Battye}, {Benabed}, {Beno{\^\i}t}, {Benoit-L{\'e}vy}, {Bernard}, {Bersanelli}, {Bielewicz}, {Bock}, {Bonaldi}, {Bonavera}, {Bond}, {Borrill}, {Bouchet}, {Boulanger}, {Bucher}, {Burigana}, {Butler}, {Calabrese}, {Cardoso}, {Catalano}, {Challinor}, {Chamballu}, {Chary}, {Chiang}, {Chluba}, {Christensen}, {Church}, {Clements}, {Colombi}, {Colombo}, {Combet}, {Coulais}, {Crill}, {Curto}, {Cuttaia}, {Danese}, {Davies}, {Davis}, {de Bernardis}, {de Rosa}, {de Zotti}, {Delabrouille}, {D{\'e}sert}, {Di Valentino}, {Dickinson}, {Diego}, {Dolag}, {Dole}, {Donzelli}, {Dor{\'e}}, {Douspis}, {Ducout}, {Dunkley}, {Dupac}, {Efstathiou}, {Elsner}, {En{\ss}lin}, {Eriksen}, {Farhang}, {Fergusson}, {Finelli}, {Forni}, {Frailis}, {Fraisse}, {Franceschi}, {Frejsel}, {Galeotta}, {Galli}, {Ganga}, {Gauthier}, {Gerbino}, {Ghosh}, {Giard},
  {Giraud-H{\'e}raud}, {Giusarma}, {Gjerl{\o}w}, {Gonz{\'a}lez-Nuevo}, {G{\'o}rski}, {Gratton}, {Gregorio}, {Gruppuso}, {Gudmundsson}, {Hamann}, {Hansen}, {Hanson}, {Harrison}, {Helou}, {Henrot-Versill{\'e}}, {Hern{\'a}ndez-Monteagudo}, {Herranz}, {Hildebrandt}, {Hivon}, {Hobson}, {Holmes}, {Hornstrup}, {Hovest}, {Huang}, {Huffenberger}, {Hurier}, {Jaffe}, {Jaffe}, {Jones}, {Juvela}, {Keih{\"a}nen}, {Keskitalo}, {Kisner}, {Kneissl}, {Knoche}, {Knox}, {Kunz}, {Kurki-Suonio}, {Lagache}, {L{\"a}hteenm{\"a}ki}, {Lamarre}, {Lasenby}, {Lattanzi}, {Lawrence}, {Leahy}, {Leonardi}, {Lesgourgues}, {Levrier}, {Lewis}, {Liguori}, {Lilje}, {Linden-V{\o}rnle}, {L{\'o}pez-Caniego}, {Lubin}, {Mac{\'\i}as-P{\'e}rez}, {Maggio}, {Maino}, {Mandolesi}, {Mangilli}, {Marchini}, {Maris}, {Martin}, {Martinelli}, {Mart{\'\i}nez-Gonz{\'a}lez}, {Masi}, {Matarrese}, {McGehee}, {Meinhold}, {Melchiorri}, {Melin}, {Mendes}, {Mennella}, {Migliaccio}, {Millea}, {Mitra}, {Miville-Desch{\^e}nes}, {Moneti}, {Montier}, {Morgante}, {Mortlock},
  {Moss}, {Munshi}, {Murphy}, {Naselsky}, {Nati}, {Natoli}, {Netterfield}, {N{\o}rgaard-Nielsen}, {Noviello}, {Novikov}, {Novikov}, {Oxborrow}, {Paci}, {Pagano}, {Pajot}, {Paladini}, {Paoletti}, {Partridge}, {Pasian}, {Patanchon}, {Pearson}, {Perdereau}, {Perotto}, {Perrotta}, {Pettorino}, {Piacentini}, {Piat}, {Pierpaoli}, {Pietrobon}, {Plaszczynski}, {Pointecouteau}, {Polenta}, {Popa}, {Pratt}, \& {Pr{\'e}zeau}}]{planck_2016}
{Planck Collaboration}, {Ade}, P.~A.~R., {Aghanim}, N., {et~al.} 2016, \aap, 594, A13

\bibitem[{{Raha} {et~al.}(1991){Raha}, {Sellwood}, {James}, \& {Kahn}}]{raha_1991}
{Raha}, N., {Sellwood}, J.~A., {James}, R.~A., \& {Kahn}, F.~D. 1991, \nat, 352, 411

\bibitem[{{Renzini} \& {Peng}(2015)}]{renzini_2015}
{Renzini}, A., \& {Peng}, Y.-j. 2015, \apjl, 801, L29

\bibitem[{{Rosas-Guevara} {et~al.}(2020){Rosas-Guevara}, {Bonoli}, {Dotti}, {Zana}, {Nelson}, {Pillepich}, {Ho}, {Izquierdo-Villalba}, {Hernquist}, \& {Pakmor}}]{RosasGuevara2020}
{Rosas-Guevara}, Y., {Bonoli}, S., {Dotti}, M., {et~al.} 2020, \mnras, 491, 2547

\bibitem[{{Rosas-Guevara} {et~al.}(2022){Rosas-Guevara}, {Bonoli}, {Dotti}, {Izquierdo-Villalba}, {Lupi}, {Zana}, {Bonetti}, {Nelson}, {Springel}, {Hernquist}, \& {Vogelsberger}}]{RosasGuevara2022}
{Rosas-Guevara}, Y., {Bonoli}, S., {Dotti}, M., {et~al.} 2022, \mnras, 512, 5339

\bibitem[{{Sakamoto} {et~al.}(1999){Sakamoto}, {Okumura}, {Ishizuki}, \& {Scoville}}]{sakamoto_1999}
{Sakamoto}, K., {Okumura}, S.~K., {Ishizuki}, S., \& {Scoville}, N.~Z. 1999, \apj, 525, 691

\bibitem[{{Salo} {et~al.}(2010){Salo}, {Laurikainen}, {Buta}, \& {Knapen}}]{salo_2010}
{Salo}, H., {Laurikainen}, E., {Buta}, R., \& {Knapen}, J.~H. 2010, \apjl, 715, L56

\bibitem[{{S{\'a}nchez-Mart{\'\i}n} {et~al.}(2023){S{\'a}nchez-Mart{\'\i}n}, {Garc{\'\i}a-G{\'o}mez}, {Masdemont}, \& {Romero-G{\'o}mez}}]{sanchezmartin_2023}
{S{\'a}nchez-Mart{\'\i}n}, P., {Garc{\'\i}a-G{\'o}mez}, C., {Masdemont}, J.~J., \& {Romero-G{\'o}mez}, M. 2023, \mnras, 520, 3909

\bibitem[{{Sellwood} \& {Wilkinson}(1993)}]{sellwood_1993}
{Sellwood}, J.~A., \& {Wilkinson}, A. 1993, Reports on Progress in Physics, 56, 173

\bibitem[{{Semczuk} {et~al.}(2024){Semczuk}, {Dehnen}, {Sch{\"o}nrich}, \& {Athanassoula}}]{semczuk_2024}
{Semczuk}, M., {Dehnen}, W., {Sch{\"o}nrich}, R., \& {Athanassoula}, E. 2024, \aap, 692, A159

\bibitem[{{Seo} {et~al.}(2019){Seo}, {Kim}, {Kwak}, {Hsieh}, {Han}, \& {Hopkins}}]{seo_2019}
{Seo}, W.-Y., {Kim}, W.-T., {Kwak}, S., {et~al.} 2019, \apj, 872, 5

\bibitem[{{Sheth} {et~al.}(2000){Sheth}, {Regan}, {Vogel}, \& {Teuben}}]{sheth_2000}
{Sheth}, K., {Regan}, M.~W., {Vogel}, S.~N., \& {Teuben}, P.~J. 2000, \apj, 532, 221

\bibitem[{{Sheth} {et~al.}(2005){Sheth}, {Vogel}, {Regan}, {Thornley}, \& {Teuben}}]{sheth_2005}
{Sheth}, K., {Vogel}, S.~N., {Regan}, M.~W., {Thornley}, M.~D., \& {Teuben}, P.~J. 2005, \apj, 632, 217

\bibitem[{{Sheth} {et~al.}(2008){Sheth}, {Elmegreen}, {Elmegreen}, {Capak}, {Abraham}, {Athanassoula}, {Ellis}, {Mobasher}, {Salvato}, {Schinnerer}, {Scoville}, {Spalsbury}, {Strubbe}, {Carollo}, {Rich}, \& {West}}]{sheth_2008}
{Sheth}, K., {Elmegreen}, D.~M., {Elmegreen}, B.~G., {et~al.} 2008, \apj, 675, 1141

\bibitem[{{Shlosman}(1994)}]{combes_1994}
{Shlosman}, I., ed. 1994, {Mass-transfer induced activity in galaxies}

\bibitem[{{Simkin} {et~al.}(1980){Simkin}, {Su}, \& {Schwarz}}]{simkin_1980}
{Simkin}, S.~M., {Su}, H.~J., \& {Schwarz}, M.~P. 1980, \apj, 237, 404

\bibitem[{{Sormani} {et~al.}(2015){Sormani}, {Binney}, \& {Magorrian}}]{sormani_2015}
{Sormani}, M.~C., {Binney}, J., \& {Magorrian}, J. 2015, \mnras, 449, 2421

\bibitem[{{van de Ven} \& {Fathi}(2010)}]{vandeven_2010ApJ...723..767V}
{van de Ven}, G., \& {Fathi}, K. 2010, \apj, 723, 767

\bibitem[{{van der Marel} \& {Cioni}(2001)}]{vandermarel_2001}
{van der Marel}, R.~P., \& {Cioni}, M.-R.~L. 2001, \aj, 122, 1807

\bibitem[{{Wang} \& {Zhou}(2025)}]{wang_2025}
{Wang}, W., \& {Zhou}, Z. 2025, \apj, 982, 129

\bibitem[{{Waters} {et~al.}(2024){Waters}, {Peterson}, {Emami}, {Shen}, {Hernquist}, {Smith}, {Vogelsberger}, {Alcock}, {Tremblay}, {Liska}, {Forbes}, \& {Moreno}}]{waters_2024}
{Waters}, T.~K., {Peterson}, C., {Emami}, R., {et~al.} 2024, \apj, 961, 193

\bibitem[{{Yu} {et~al.}(2018){Yu}, {Ho}, {Barth}, \& {Li}}]{yu_2018}
{Yu}, S.-Y., {Ho}, L.~C., {Barth}, A.~J., \& {Li}, Z.-Y. 2018, \apj, 862, 13

\bibitem[{{Zana} {et~al.}(2022){Zana}, {Lupi}, {Bonetti}, {Dotti}, {Rosas-Guevara}, {Izquierdo-Villalba}, {Bonoli}, {Hernquist}, \& {Nelson}}]{zana_2022}
{Zana}, T., {Lupi}, A., {Bonetti}, M., {et~al.} 2022, \mnras, 515, 1524

\bibitem[{{Zee} {et~al.}(2023){Zee}, {Paudel}, {Moon}, \& {Yoon}}]{zee_2023}
{Zee}, W.-B.~G., {Paudel}, S., {Moon}, J.-S., \& {Yoon}, S.-J. 2023, \apj, 949, 91

\bibitem[{{Zhao} {et~al.}(2020){Zhao}, {Du}, {Ho}, {Debattista}, \& {Shi}}]{Zhao2020}
{Zhao}, D., {Du}, M., {Ho}, L.~C., {Debattista}, V.~P., \& {Shi}, J. 2020, \apj, 904, 170

\bibitem[{{Zhou} {et~al.}(2020){Zhou}, {Zhu}, {Wang}, \& {Feng}}]{zhou_2020}
{Zhou}, Z.-B., {Zhu}, W., {Wang}, Y., \& {Feng}, L.-L. 2020, \apj, 895, 92

\end{thebibliography}

\end{document}